\documentclass[aps, prb, twocolumn, amssymb, amsmath, showpacs, superscriptaddress]{revtex4-1}

\usepackage{bm}
\usepackage{times}
\usepackage{graphicx}
\usepackage{color}
\usepackage{dcolumn}
\usepackage[colorlinks=true, letterpaper=true, pdfstartview=FitV, linkcolor=blue, citecolor=blue, urlcolor=blue]{hyperref}
\usepackage{appendix}
\usepackage[normalem]{ulem}

\begin{document}
	
	\title{Spin-order dependent anomalous Hall effect and magneto-optical effect in noncollinear antiferromagnets Mn$_{3}X$N ($X$ = Ga, Zn, Ag, and Ni)}
	
	\author{Xiaodong Zhou}
	\affiliation {Key Laboratory of Advanced Optoelectronic Quantum Architecture and Measurement, Ministry of Education, School of Physics, Beijing Institute of Technology, Beijing 100081, China}
	
	\author{Jan-Philipp Hanke}
	\affiliation{Institute of Physics, Johannes Gutenberg University Mainz, 55099 Mainz, Germany}
	
	\author{Wanxiang Feng}
	\thanks{wxfeng@bit.edu.cn}
	\affiliation {Key Laboratory of Advanced Optoelectronic Quantum Architecture and Measurement, Ministry of Education, School of Physics, Beijing Institute of Technology, Beijing 100081, China}
	\affiliation{Peter Gr\"unberg Institut and Institute for Advanced Simulation, Forschungszentrum J\"ulich and JARA, 52425 J\"ulich, Germany}
	
	\author{Fei Li}
	\affiliation {Key Laboratory of Advanced Optoelectronic Quantum Architecture and Measurement, Ministry of Education, School of Physics, Beijing Institute of Technology, Beijing 100081, China}
	
	\author{Guang-Yu Guo}
	\affiliation{Department of Physics and Center for Theoretical Physics, National Taiwan University, Taipei 10617, Taiwan}
	\affiliation{Physics Division, National Center for Theoretical Sciences, Hsinchu 30013, Taiwan}
	
	\author{Yugui Yao}
	\affiliation {Key Laboratory of Advanced Optoelectronic Quantum Architecture and Measurement, Ministry of Education, School of Physics, Beijing Institute of Technology, Beijing 100081, China}
	
	\author{Stefan Bl\"ugel}
	\affiliation{Peter Gr\"unberg Institut and Institute for Advanced Simulation, Forschungszentrum J\"ulich and JARA, 52425 J\"ulich, Germany}
	
	\author{Yuriy Mokrousov}
	\affiliation{Peter Gr\"unberg Institut and Institute for Advanced Simulation, Forschungszentrum J\"ulich and JARA, 52425 J\"ulich, Germany}
	\affiliation{Institute of Physics, Johannes Gutenberg University Mainz, 55099 Mainz, Germany}
	
	\date{\today}

	\begin{abstract}
		The anomalous Hall effect (AHE) and the magneto-optical effect (MOE) are two prominent manifestations of time-reversal symmetry breaking in magnetic materials. Noncollinear  antiferromagnets (AFMs) have recently attracted a lot of attention owing to the potential emergence of exotic spin orders on geometrically frustrated lattices, which can be characterized by corresponding spin chiralities.  By performing first-principles density functional calculations together with group-theory analysis and tight-binding modelling, here we systematically study the spin-order dependent AHE and MOE in representative noncollinear AFMs Mn$_{3}X$N ($X$ = Ga, Zn, Ag, and Ni).  The symmetry-related tensor shape of the intrinsic anomalous Hall conductivity (IAHC) for different spin orders is determined by analyzing the relevant magnetic point groups.  We show that while only the ${xy}$ component of the IAHC tensor is nonzero for right-handed spin chirality, all other elements $-$ $\sigma_{xy}$, $\sigma_{yz}$, and $\sigma_{zx}$ $-$ are nonvanishing for a state with left-handed spin chirality owing to lowering of the symmetry.  Our tight-binding arguments reveal that the magnitude of IAHC relies on the details of the band structure and that $\sigma_{xy}$ is periodically modulated as the spin rotates in-plane. The IAHC obtained from first principles is found to be rather large, e.g., it amounts to 359 S/cm in Mn$_{3}$AgN, which is comparable to other well-known noncollinear AFMs such as Mn$_{3}$Ir and Mn$_{3}$Ge.  We evaluate also the magnetic anisotropy energy and find that the evolution of spin order is related to the number of valence electrons in the $X$ ion. Interestingly, the left-handed spin chirality could exist in Mn$_{3}X$N with some particular spin configurations.  By extending our analysis to finite frequencies, we calculate the optical isotropy [$\sigma_{xx}(\omega)\approx\sigma_{yy}(\omega)\approx\sigma_{zz}(\omega)$] and the magneto-optical anisotropy [$\sigma_{xy}(\omega)\neq\sigma_{yz}(\omega)\neq\sigma_{zx}(\omega)$] of Mn$_{3}X$N.  Similar to the IAHC, the magneto-optical Kerr and Faraday spectra depend strongly on the spin order.  The Kerr rotation angles in Mn$_3X$N are in the range of $0.3\sim0.4$ deg, which is large and comparable to other noncollinear AFMs like Mn$_{3}$Pt and Mn$_{3}$Sn. Our finding of large AHE and MOE in Mn$_3X$N suggests that these materials present an excellent antiferromagnetic platform for realizing novel spintronics and magneto-optical devices.  We argue that the spin-order dependent AHE and MOE are indispensable in detecting complex spin structures in noncollinear AFMs.
	\end{abstract}
	
	\maketitle

	\section{Introduction}\label{sec1}
	
	The anomalous Hall effect (AHE)~\cite{Nagaosa2010} and the magneto-optical effect (MOE)~\cite{Ebert1996,Oppeneer2001,Antonov2004} are fundamental phenomena in condensed matter physics and they have become appealing techniques to detect and measure magnetism by electric and optical means, respectively.  Usually occuring in ferromagnetic metals, the AHE is characterized by a transverse voltage drop resulting from a longitudinal charge current in the absence of applied magnetic fields. There are two distinct contributions to the AHE, that is, the extrinsic one~\cite{Smit1955,Smit1958,Berger1970} depending on scattering of electron off impurities or due to disorder, and the intrinsic one~\cite{Sundaram1999,YG-Yao2004} solely determined by the Berry phase effect~\cite{D-Xiao2010} in a pristine crystal. Both of these mechanisms originate from time-reversal ($T$) symmetry breaking in combination with spin-orbit coupling (SOC)~\cite{Nagaosa2010}.  The intrinsic AHE can be accurately calculated from electronic-structure theory on the \textit{ab initio} level, and examples include studies of Fe~\cite{YG-Yao2004,XJ-Wang2006}, Co~\cite{XJ-Wang2007,Roman2009}, SrRuO$_{3}$~\cite{Fang2003,Mathieu2004}, Mn$_{5}$Ge$_{3}$~\cite{CG-Zeng2006}, and CuCr$_{2}$Se$_{4-x}$Br$_{x}$~\cite{YG-Yao2007}.  Referring to the Kubo formula~\cite{Kubo1957,CS-Wang1974}, the intrinsic anomalous Hall conductivity (IAHC) can be straightforwardly extended to the ac case (as given by the optical Hall conductivity), which is intimately related to the magneto-optical Kerr and Faraday effects (MOKE and MOFE) [see Eqs.~\eqref{eq:kerr} and~\eqref{eq:faraday} below].  Phenomenally, the MOKE and MOFE refer to the rotation of the polarization plane when a linearly polarized light is reflected from, or transmitted through a magnetic material, respectively.  Owing to their similar physical nature, the intrinsic AHE is often studied together with MOKE and MOFE.
	
	\begin{figure*}
		\includegraphics[width=2\columnwidth]{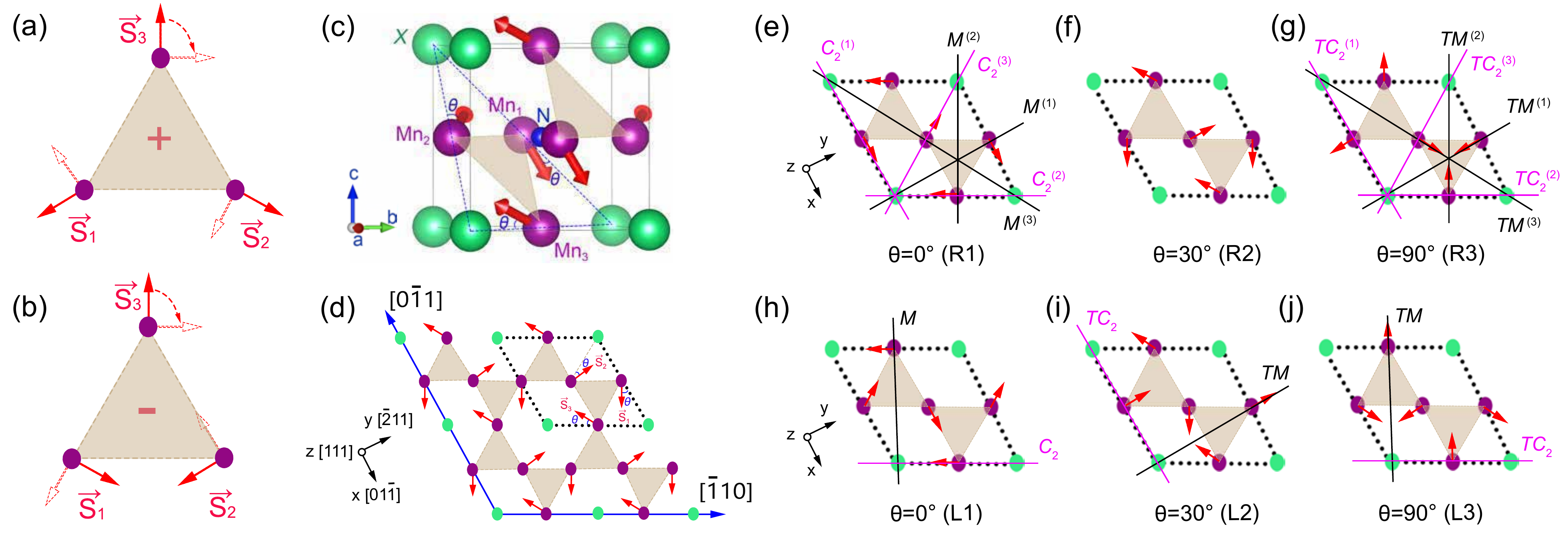}
		\caption{(Color online) (a) Right-handed ($\kappa=+1$) and (b) left-handed ($\kappa=-1$) vector spin chiralities in coplanar noncollinear spin systems.  The open arrows indicate the clockwise rotation of spin with a uniform angle, which results in a different spin configuration with the same spin chirality.  (c) The crystal and magnetic structures of Mn$_{3}X$N ($X$ = Ga, Zn, Ag, and Ni).  The purple, green, and blue balls represent Mn, $X$, and N atoms, respectively.  The spin magnetic moments originate mainly from  Mn atoms, while the spin polarization of $X$ and N atoms is negligible.  The spins on three Mn-sublattices (Mn$_{1}$, Mn$_{2}$, and Mn$_{3}$) are indicated by red arrows that are aligned within the (111) plane (here, the right-handed spin chirality is shown as an example).  The angles between neighboring spins are always 120$^{\circ}$, while the spins can simultaneously rotate within the (111) plane that is characterized by an azimuthal angle $\theta$ away from the diagonals of the face.  (d) The (111) plane of Mn$_{3}X$N, which can be regarded as a kagome lattice of Mn atoms.  The dotted lines mark the two-dimensional unit cell.  (e)-(g) The R1, R2, and R3 phases with the right-handed spin chirality.   There are one three-fold rotation axis ($C_{3}$, which is along the $[111]$ direction ($z$ axis)), three two-fold rotation axes ($C_{2}^{(1)}$, $C_{2}^{(2)}$, and $C_{2}^{(3)}$), and three mirror planes ($M^{(1)}$, $M^{(2)}$, and $M^{(3)}$) in the R1 phase; only $C_{3}$ axis is preserved in the R2 phase; the time-reversal symmetry $T$ has to be combined with a two-fold rotation and mirror symmetries in the R3 phase.  (h)-(j) The L1, L2, and L3 phases with the left-handed spin chirality.  There are one two-fold rotation axis ($C_{2}$) and one mirror plane ($M$) in the L1 phase; the time-reversal symmetry $T$ is combined with two-fold rotation and mirror symmetries in both the L2 and the L3 phases.}
		\label{fig1}
	\end{figure*}
	
	As the AHE and the MOE are commonly considered to be proportional to the magnetization, most of the materials studied to date with respect to these phenomena are ferromagnets (FMs) and ferrimagnets (FiMs), while antiferromagnets (AFMs) are naively expected to have neither AHE nor MOE due to their vanishing net magnetization. Although $T$ symmetry is broken in AFMs, its combination $TS$ with other spatial symmetries $S$ (e.g., fractional translations or inversion) can reinstate Kramers theorem such that AHE and MOE vanish. A simple example is the one-dimensional collinear bipartite antiferromagnet~\cite{Herring1966}, where $S$ is the fractional translation by half of the vector connecting the two sublattices.  Another example is the two-dimensional honeycomb lattice with collinear N{\'e}el order (as realized, e.g., in the bilayer MnPSe$_{3}$)~\cite{Sivadas2016}, which has natively the combined symmetry $TI$ although time-reversal symmetry $T$ and spatial inversion symmetry $I$ are both broken individually. The application of an electric field perpendicular to the film plane will manifest in broken $TI$ symmetry and band exchange splitting that generates the MOKE~\cite{Sivadas2016}. Such electrically driven  MOKE has been realized, e.g., in multiferroic Cr-based metallorganic perovskites~\cite{FR-Fan2017}.  Therefore, the AHE and the MOE, as the most fundamental fingerprints of $T$ symmetry breaking in matter, can in principle exist in AFMs if certain crystal symmetries are absent, even though the net magnetization vanishes. Notably, the cluster multipole theory proposed by Suzuki \textit{et al.}~\cite{Suzuki2017,Suzuki2018} has been recently applied to interpret the origin of AHE in AFMs.
	
	Leaving aside collinear AFMs, recent works~\cite{Ohgushi2000,Shindou2001,Hanke2017,Shiomi2018,J-Zhou2016,WX-Feng2018,H-Chen2014,Kubler2014,GY-Guo2017,Y-Zhang2017,Nakatsuji2015,Nayak2016,Kiyohara2016,Ikhlas2017,WX-Feng2015,Higo2018} revealed that noncollinear AFMs can also host nonvanishing AHE and MOE.  Two types  of noncollinear AFMs can be considered: noncoplanar and coplanar, which are characterized by scalar and vector spin chiralities, respectively~\cite{Kawamura2001}.  On the one hand, the nonzero scalar spin chirality $\chi=\boldsymbol{S}_{i}\cdot(\boldsymbol{S}_{j}\times\boldsymbol{S}_{k})$  (where $\boldsymbol{S}_{i}$, $\boldsymbol{S}_{j}$, and $\boldsymbol{S}_{k}$ denote three neighboring noncoplanar spins) will generate a fictitious magnetic field that makes the electrons feel a real-space Berry phase while hopping in the spin lattice~\cite{Ohgushi2000,Shindou2001}.  Consequently, the AHE can emerge in noncoplanar AFMs without SOC, which is referred to the topological Hall effect that has been theoretically predicted~\cite{Shindou2001,Hanke2017} and experimentally observed~\cite{Shiomi2018}, for instance, in disordered $\gamma$-Fe$_{x}$Mn$_{1-x}$ alloys.  Moreover, the quantized version of the topological Hall effect was reported in the layered noncoplanar noncollinear K$_{0.5}$RhO$_{2}$ AFM insulator~\cite{J-Zhou2016}.  Extending these findings, Feng \textit{et al.}~\cite{WX-Feng2018} proposed that topological MOE and quantum topological MOE exist in $\gamma$-Fe$_{x}$Mn$_{1-x}$ and K$_{0.5}$RhO$_{2}$, respectively.
	
	Instead of the scalar spin chirality (which vanishes for coplanar spin configurations), the finite vector spin chirality~\cite{Kawamura2001},
	\begin{equation}\label{eq:kappa}
	\kappa=\frac{2}{3\sqrt{3}}\sum_{\langle ij\rangle}\left[\boldsymbol{S}_{i}\times\boldsymbol{S}_{j}\right]_{z},
	\end{equation}
	where $\langle ij\rangle$ runs over the nearest neighboring spins, is an important quantity in coplanar noncollinear AFMs such as cubic Mn$_{3}X$ ($X$ = Rh, Ir, Pt) and hexagonal Mn$_{3}Y$ ($Y$ = Ge, Sn, Ga).  The Mn atoms in the (111) plane of Mn$_{3}X$ and in the (0001) plane of Mn$_{3}Y$ are arranged into a kagome lattice, while Mn$_{3}X$ and Mn$_{3}Y$ have opposite vector spin chiralities~\cite{Y-Zhang2017} with $\kappa=+1$ (right-handed state) and $\kappa=-1$ (left-handed state) [see Figs.~\ref{fig1}(a) and~\ref{fig1}(b)], respectively.  The concept of right- and left-handed states adopted here follows the convention of Ref.~\onlinecite{Kawamura2001}.  For both right- and left-handed spin chiralities, the spins can be simultaneously rotated within the plane, further resulting in different spin configurations [see Figs.~\ref{fig1}(a) and~\ref{fig1}(b)], e.g., the T1 and the T2 phases in Mn$_{3}X$~\cite{WX-Feng2015} as well as the type-A and the type-B phases in Mn$_{3}Y$~\cite{GY-Guo2017}.  The vector spin chirality and the spin rotation discussed here allow us to characterize coplanar AFMs that have a 120$^\circ$ noncollinear magnetic ordering.  For the AHE, Chen \textit{et al.}~\cite{H-Chen2014} discovered theoretically that Mn$_{3}$Ir has unexpectedly large IAHC and several other groups predicted the IAHC in Mn$_{3}Y$ with comparable magnitudes~\cite{Kubler2014,GY-Guo2017,Y-Zhang2017}.  At the same time, the AHE in Mn$_{3}Y$ has been experimentally confirmed~\cite{Nakatsuji2015,Nayak2016,Kiyohara2016,Ikhlas2017}.  Because of the close relationship to AHE, Feng \textit{et al.}~\cite{WX-Feng2015} first predicted that large MOKE can emerge in Mn$_{3}X$ even though the net magnetization is zero.  Eventually, Higo \textit{et al.}~\cite{Higo2018} successfully measured large zero-field Kerr rotation angles in Mn$_{3}$Sn at room temperature.
	
	
	In addition to Mn$_3X$ and Mn$_3Y$, the antiperovskite Mn$_3X$N ($X$ = Ga, Zn, Ag, Ni, etc.) is another important class of coplanar noncollinear AFMs~\cite{Singh2018}, which was known since the 1970s~\cite{Bertaut1968,Fruchart1978}.  Compared to Mn$_3X$, the $X$ atoms in Mn$_3X$N also occupy the corners of the cube [see Fig.~\ref{fig1}(c)] and the face-centered Mn atoms are arranged into a kagome lattice in the (111) plane [see Fig.~\ref{fig1}(d)], while there is an additional N atom located in the center of the cube [see Fig.~\ref{fig1}(c)].  Despite the structural similarity, some unique physical properties have been found in Mn$_3X$N, such as magnetovolume effects~\cite{Gomonaj1989,Gomonaj1992,WS-Kim2003,Lukashev2008,Lukashev2010,Takenaka2014,SH-Deng2015,Zemen2017a} and magnetocaloric effects~\cite{Y-Sun2012,Matsunami2014,KW-Shi2016,Zemen2017} that stem from a strong coupling between spin, lattice, and heat.  The most interesting discovery in Mn$_3X$N may be the giant negative thermal expansion that was observed in the first-order phase transition from a paramagnetic state to a noncollinear antiferromagnetic state with decreasing temperature  $\mathtt{T}$.  Below the N{\'e}el temperature ($\mathtt{T_{N}}$), a second-order phase transition between two different noncollinear antiferromagnetic states, which are featured by a nearly constant volume but the change of spin configuration, possibly occurs.  
	
	Taking Mn$_3$NiN as an example~\cite{Fruchart1978}, all the spins point along the diagonals of the face if $\mathtt{T}<$163 K (the so-called $\Gamma^{5g}$ configuration), while in the temperature range of 163 K $<\mathtt{T}<$ 266 K the spins can point to the center of the triangle formed by three nearest-neighboring Mn atoms (the so-called $\Gamma^{4g}$ configuration).  The $\Gamma^{5g}$ and the $\Gamma^{4g}$ spin configurations are named as R1 ($\theta=0^{\circ}$) and R3 ($\theta=90^{\circ}$) phases in this work [see Figs.~\ref{fig1}(e) and~\ref{fig1}(g), where the azimuthal angle $\theta$ measures the rotation of the spins starting from the diagonals of the face], respectively.  An intermediate state ($0^{\circ}<\theta<90^{\circ}$) between the R1 and R3 phases, referred to as the R2 phase (see Fig.~\ref{fig1}(f) with $\theta=30^{\circ}$ as an example), was proposed to exist~\cite{Gomonaj1989,Gomonaj1992}.  Such nontrivial magnetic orders are also believed to occur in other Mn$_3X$N compounds~\cite{Bertaut1968,Fruchart1978,Gomonaj1989,Gomonaj1992}, as recently clarified by Mochizuki \textit{et al.}~\cite{Mochizuki2018} using a classical spin model together with the replica-exchange Monte Carlo simulation.  However, the details of the changes in spin configurations from R1 phase, passing through the R2 phase to R3 phase, and how they affect the relevant physical properties (e.g., AHE and MOE) are still unclear.  Moreover, although only the right-handed spin chirality was reported in the previous literature, the left-handed spin chirality as a counterpart [Fig.~\ref{fig1}(h-j)] could also exist, e.g., in Mn$_{3}$NiN, because of the favorable total energy for a particular $\theta$ [see Fig.~\ref{fig4}(a)].
	
	In this work, using  first-principles density functional theory together with group-theory analysis and tight-binding modelling, we systematically investigate the effect of \textit{spin order} on the intrinsic AHE as well as the MOKE and the MOFE in coplanar noncollinear AFMs Mn$_{3}X$N ($X$ = Ga, Zn, Ag, and Ni).  The \textit{spin order} considered here has dual implications, i.e., spin chiralities (right- and left-handed states) and spin configurations [regarding the different spin orientations by simultaneously rotating the spins within the (111) plane].  In Sec.~\ref{sec2}, we first identify the antisymmetric shape of the IAHC tensor (i.e., zero and nonzero elements) for different spin orders by a group theoretical analysis.  For the right-handed spin chirality, only $\sigma_{xy}$ is nonzero (except for two particular spin configurations: $\theta=0^{\circ}$ and $180^{\circ}$); for the left-handed spin chirality, all three off-diagonal elements ($\sigma_{xy}$, $\sigma_{yz}$, and $\sigma_{zx}$) can be nonzero (except for some particular spin configurations, e.g., $\theta=0^{\circ}$ and $60^{\circ}$ for $\sigma_{xy}$, $\theta=30^{\circ}$ and $210^{\circ}$ for $\sigma_{yz}$, $\theta=120^{\circ}$ and $300^{\circ}$ for $\sigma_{zx}$).  The results of the group-theory analysis are further confirmed by both tight-binding modelling (Sec.~\ref{sec3}) and first-principles calculations (Sec.~\ref{sec4-1}).  In addition to the IAHC, the magnetic anisotropy energy (MAE) has also been accessed and the in-plane easy spin orientation is determined (Sec.~\ref{sec4-1}).  
	
	Considering Mn$_{3}$NiN as a prototype, we extend the study of IAHC to the optical Hall conductivity [$\sigma_{xy}(\omega)$, $\sigma_{yz}(\omega)$ $\sigma_{zx}(\omega)$] as well as the corresponding diagonal elements [$\sigma_{xx}(\omega)$, $\sigma_{yy}(\omega)$, and $\sigma_{zz}(\omega)$] (Sec.~\ref{sec4-2}).  The spin order hardly affects the diagonal elements, whereas a significant dependence on the spin order is observed in the off-diagonal elements akin to the IAHC.  Subsequently in Sec.~\ref{sec4-3}, the MOKE and the MOFE are computed from the optical conductivity for all Mn$_{3}X$N ($X$ = Ga, Zn, Ag, and Ni). Kerr and Faraday spectra exhibit a distinct dependence on the spin order, which they inherit from the optical Hall conductivity.  The computed Kerr and Faraday rotation angles in Mn$_{3}X$N are comparable to the ones in Mn$_{3}X$ studied in our previous work~\cite{WX-Feng2015}.  The magneto-optical anisotropy, originating from the nonequivalent off-diagonal elements of optical conductivity, is explored for both right- and left-handed spin chiralities.  Finally, the summary is drawn in Sec.~\ref{sec5}.  Our work reveals that the AHE and the MOE depend strongly on the spin order in noncollinear AFMs Mn$_{3}X$N which suggests that complex noncollinear spin structures can be uniquely classified in experiments by measuring AHE and MOE.

	\begin{table*}[htpb]
		\caption{The magnetic space and point groups as well as the nonzero elements of IAHC for Mn$_{3}X$N for different spin orders characterized by the azimuthal angle $\theta$ and the vector spin chirality $\kappa$.  The magnetic space and point groups exhibit a period of $\pi$ ($\pi/3$) in $\theta$ for right-handed (left-handed) spin chirality.  The IAHC is considered as a pseudovector, i.e., $\boldsymbol{\sigma}=[\sigma^{x},\sigma^{y},\sigma^{z}]=[\sigma_{yz},\sigma_{zx},\sigma_{xy}]$, which is expressed in the Cartesian coordinate system defined in Fig.~\ref{fig1}.  The nonzero elements of IAHC are in complete accord with the tight-binding and first-principles calculations, shown in Figs.~\ref{fig2}(c), ~\ref{fig4}(b), and~\ref{fig4}(c), respectively.}
		\label{tab1}
		\begin{ruledtabular}
			\begingroup
			\setlength{\tabcolsep}{4.5pt} 
			\renewcommand{\arraystretch}{1.5} 
			\begin{tabular}{lccccccccccccccc}
				
				\multicolumn{2}{c}{} &
				\multicolumn{13}{c}{azimuthal angle $\theta$} &  \\
				\cline{3-15}
				
				&$\kappa$&$0^{\circ}$&$15^{\circ}$&$30^{\circ}$&$45^{\circ}$&$60^{\circ}$&$75^{\circ}$&$90^{\circ}$&$105^{\circ}$&$120^{\circ}$&$135^{\circ}$&$150^{\circ}$&$165^{\circ}$&$180^{\circ}$&\\
				
				\hline
				
				magnetic space group & $+1$ & $R\bar{3}m$ & $R\bar{3}$ & $R\bar{3}$ & $R\bar{3}$ & $R\bar{3}$ & $R\bar{3}$ & $R\bar{3}m^{\prime}$ & $R\bar{3}$ & $R\bar{3}$ & $R\bar{3}$ & $R\bar{3}$ & $R\bar{3}$ & $R\bar{3}m$ \\
				
				& $-1$  & $C2/m$ & $P\bar{1}$ & $C2^{\prime}/m^{\prime}$ & $P\bar{1}$ & $C2/m$ & $P\bar{1}$ & $C2^{\prime}/m^{\prime}$ & $P\bar{1}$ & $C2/m$ & $P\bar{1}$ & $C2^{\prime}/m^{\prime}$ & $P\bar{1}$ & $C2/m$ \\
				
				\hline
				
				magnetic point group & $+1$ & $\bar{3}1m$ & $\bar{3}$ & $\bar{3}$ & $\bar{3}$ & $\bar{3}$ & $\bar{3}$ & $\bar{3}1m^{\prime}$ & $\bar{3}$ & $\bar{3}$ & $\bar{3}$ & $\bar{3}$ & $\bar{3}$ & $\bar{3}1m$ \\
				
				& $-1$  & $2/m$ & $\bar{1}$ & $2^{\prime}/m^{\prime}$ & $\bar{1}$ & $2/m$ & $\bar{1}$ & $2^{\prime}/m^{\prime}$ & $\bar{1}$ & $2/m$ & $\bar{1}$ & $2^{\prime}/m^{\prime}$ & $\bar{1}$& $2/m$ \\
				
				\hline
				
				nonzero elements & $+1$ & -- & $\sigma_{xy}$ & $\sigma_{xy}$ & $\sigma_{xy}$ & $\sigma_{xy}$ & $\sigma_{xy}$ & $\sigma_{xy}$ & $\sigma_{xy}$ & $\sigma_{xy}$ & $\sigma_{xy}$ & $\sigma_{xy}$ & $\sigma_{xy}$ & -- \\
				
				of IAHC & $-1$ & \vtop{\hbox{\strut $\;\:$--}\hbox{\strut $\sigma_{yz}$}\hbox{\strut $\sigma_{zx}$}} & \vtop{\hbox{\strut $\sigma_{xy}$}\hbox{\strut $\sigma_{yz}$}\hbox{\strut $\sigma_{zx}$}} &  \vtop{\hbox{\strut $\sigma_{xy}$}\hbox{\strut $\;\:$--}\hbox{\strut $\sigma_{zx}$}}  & \vtop{\hbox{\strut $\sigma_{xy}$}\hbox{\strut $\sigma_{yz}$}\hbox{\strut $\sigma_{zx}$}} & \vtop{\hbox{\strut $\;\:$--}\hbox{\strut $\sigma_{yz}$}\hbox{\strut $\sigma_{zx}$}} & \vtop{\hbox{\strut $\sigma_{xy}$}\hbox{\strut $\sigma_{yz}$}\hbox{\strut $\sigma_{zx}$}} & \vtop{\hbox{\strut $\sigma_{xy}$}\hbox{\strut $\sigma_{yz}$}\hbox{\strut $\sigma_{zx}$}} & \vtop{\hbox{\strut $\sigma_{xy}$}\hbox{\strut $\sigma_{yz}$}\hbox{\strut $\sigma_{zx}$}} & \vtop{\hbox{\strut $\;\:$--}\hbox{\strut $\sigma_{yz}$}\hbox{\strut $\;\:$--}}  & \vtop{\hbox{\strut $\sigma_{xy}$}\hbox{\strut $\sigma_{yz}$}\hbox{\strut $\sigma_{zx}$}} & \vtop{\hbox{\strut $\sigma_{xy}$}\hbox{\strut $\sigma_{yz}$}\hbox{\strut $\sigma_{zx}$}}  & \vtop{\hbox{\strut $\sigma_{xy}$}\hbox{\strut $\sigma_{yz}$}\hbox{\strut $\sigma_{zx}$}} & \vtop{\hbox{\strut $\;\:$--}\hbox{\strut $\sigma_{yz}$}\hbox{\strut $\sigma_{zx}$}} 
				
			\end{tabular}
			\endgroup
		\end{ruledtabular}
	\end{table*}

	\section{Group theory analysis}\label{sec2}
	
	In this section, we determine the magnetic space and point groups of Mn$_{3}X$N for given spin orders, and then identify the nonzero elements of IAHC from group theory.  The magnetic groups computed with the \textsc{isotropy} code~\cite{Stokes,Stokes2005} are listed in Table~\ref{tab1}, from which one can observe that the magnetic groups vary in the azimuthal angle $\theta$ with a period of $\pi$ for right-handed spin chirality, but with a period of $\pi/3$ for left-handed spin chirality. This indicates that the magnetic groups that need to be analyzed are limited to a finite number.  Furthermore, it is sufficient to restrict the analysis to magnetic point groups since the IAHC~\cite{Kubo1957,CS-Wang1974,YG-Yao2004},
	\begin{equation}\label{eq:IAHC}
	\sigma_{\alpha\beta} = -\dfrac{e^{2}}{\hbar}\int_{BZ}\frac{d^{3}k}{(2\pi)^{3}}\Omega_{\alpha\beta}(\bm{k}),
	\end{equation}
	is translationally invariant. In the above expression $\Omega_{\alpha\beta}(\bm{k})=\sum_{n}f_{n}(\bm{k})\Omega_{n,\alpha\beta}(\bm{k})$ is the momentum-space Berry curvature, with the Fermi-Dirac distribution function $f_{n}(\bm{k})$ and the band-resolved Berry curvature
	\begin{equation}\label{eq:BerryCur}
	\Omega_{n,\alpha\beta}\left(\bm{k}\right) = -2 \mathrm{Im}\sum_{n^{\prime} \neq n}\frac{\left\langle \psi_{n\bm{k}}\right|\hat{v}_{\alpha}\left| \psi_{n^{\prime}\bm{k}} \right\rangle \left\langle \psi_{n^{\prime}\bm{k}}\right|\hat{v}_{\beta}\left|\psi_{n\bm{k}} \right\rangle}{\left(\omega_{n^{\prime}\bm{k}}-\omega_{n\bm{k}}\right)^{2}}.
	\end{equation}
	Here $\hat{v}_{\alpha}$ is the velocity operator along the $\alpha$th Cartesian direction, and $\psi_{n\bm{k}}$ ($\hbar\omega_{n\bm{k}}=\epsilon_{n\bm{k}}$) is the eigenvector (eigenvalue) to the band index $n$ and the momentum $\bm{k}$.   Since the IAHC and the Berry curvature can be regarded as pseudovectors, just like spin, their vector-form notations $\boldsymbol{\sigma}=[\sigma^{x},\sigma^{y},\sigma^{z}]=[\sigma_{yz},\sigma_{zx},\sigma_{xy}]$ and $\boldsymbol{\Omega}_{n}=[\Omega_{n}^{x},\Omega_{n}^{y},\Omega_{n}^{z}]=[\Omega_{n,yz},\Omega_{n,zx},\Omega_{n,xy}]$ are used here for convenience.
	
	Let us start with the right-handed spin chirality by considering the three non-repetitive magnetic point groups: $\bar{3}1m$ [$\theta=n\pi$], $\bar{3}1m^{\prime}$ [$\theta=(n +\frac{1}{2})\pi$], and $\bar{3}$ [$\theta \neq n\pi \text{ and }\theta \neq (n +\frac{1}{2})\pi$] with $n\in\mathbb{N}$ (see Tab.~\ref{tab1}).  First, $\bar{3}1m$ belongs to the type-I magnetic point group, i.e., it is identical to the crystallographic point group $D_{3d}$.  As seen from Fig.~\ref{fig1}(e), it has one three-fold rotation axis ($C_{3}$), three two-fold rotation axes ($C_{2}^{(1)}$, $C_{2}^{(2)}$, and $C_{2}^{(3)}$) and three mirror planes ($M^{(1)}$, $M^{(2)}$, and $M^{(3)}$).  As mentioned before, $\boldsymbol{\Omega}_{n}$ is a pseudovector, and the mirror operation $M^{(1)}$ (parallel to the $yz$ plane) changes the sign of $\Omega_{n}^{y}$ and $\Omega_{n}^{z}$, but preserves $\Omega_{n}^{x}$.  This indicates that $\Omega_{n}^{y}$ and $\Omega_{n}^{z}$ are odd functions along the $k_{x}$ direction in momentum space, while $\Omega_{n}^{x}$ is an even function.  Correspondingly,  integrating the Berry curvature over the entire Brillouin zone should give  $\boldsymbol{\sigma}=[\sigma^{x},0,0]$.  The role of $C_{2}^{(1)}$ is the same as that of $M^{(1)}$.   The other two mirror (two-fold rotation) symmetries are related to $M^{(1)}$ ($C_{2}^{(1)}$) by the $C_{3}$ rotation, which transforms $[\sigma^{x},0,0]$ into $[-\frac{1}{2}\sigma^{x},-\frac{\sqrt{3}}{2}\sigma^{x},0]$ and $[-\frac{1}{2}\sigma^{x},\frac{\sqrt{3}}{2}\sigma^{x},0]$.  Therefore, all components of IAHC are zero, i.e., $\boldsymbol{\sigma}=[0,0,0]$, owing to the symmetries of the group $\bar{3}1m$.  Second, $\bar{3}$ is also a type-I magnetic point group, which is identical to the crystallographic point group $C_{3i}$.  Compared to $D_{3d}$, all $C_{2}$ and $M$ operations are absent whereas only the $C_{3}$ operation is left [see Fig.~\ref{fig1}(f)].  In this situation, the components of $\boldsymbol{\sigma}$ that are normal to the $C_{3}$ axis disappear due to the cancellations of $\Omega_{n}^{x}$ and $\Omega_{n}^{y}$ in the $k_{x}$--$k_{y}$ plane.  This gives rise to $\boldsymbol{\sigma}=[0,0,\sigma^{z}]$.  Finally, $\bar{3}1m^{\prime}=C_{3i} \oplus T(D_{3d}-C_{3i})$ is a type-III magnetic point group as it contains operations combining time and space symmetries.  Here, $T(D_{3d}-C_{3i})$ is the set of three $TM$ and three $TC_{2}$ operations depicted in Fig.~\ref{fig1}(g).  With respect to the mirror symmetry $M^{(1)}$, $\Omega_{n}^{x}$ is even but $\Omega_{n}^{y}$ and $\Omega_{n}^{z}$ are odd; with respect to the time-reversal symmetry $T$, all of $\Omega_{n}^{x}$, $\Omega_{n}^{y}$, and $\Omega_{n}^{z}$ are odd; hence, with respect to the $TM^{(1)}$ symmetry, $\Omega_{n}^{x}$ is odd but $\Omega_{n}^{y}$ and $\Omega_{n}^{z}$ are even, resulting in $\boldsymbol{\sigma}=[0,\sigma^{y},\sigma^{z}]$.  $TC_{2}^{(1)}$ plays the same role, just like $TM^{(1)}$ does.  The other two $TM$ ($TC_{2}$) symmetries are related to $TM^{(1)}$ ($TC_{2}^{(1)}$) by the $C_{3}$ rotation in the subgroup $C_{3i}$, which forces $\sigma^{y}$ to be zero but allows finite $\sigma^{z}$. Thus, the IAHC tensor shape is $\boldsymbol{\sigma}=[0,0,\sigma^{z}]$ in the magnetic point group $\bar{3}1m^{\prime}$.  To summarize, for the right-handed spin chirality  only $\sigma^{z}$ can be nonzero, except for $\theta=n\pi$ where all components of the IAHC vanish.
	
	Next, we turn to the left-handed spin chirality, which also has three non-repetitive magnetic point groups: $2/m$ [$\theta=n\frac{\pi}{3}$], $2^{\prime}/m^{\prime}$ [$\theta=(n +\frac{1}{2})\frac{\pi}{3}$], and $\bar{1}$ [$\theta \neq n\frac{\pi}{3} \text{ and }\theta \neq (n +\frac{1}{2})\frac{\pi}{3}$] with $n\in\mathbb{N}$ (see Tab.~\ref{tab1}).  First, $2/m$ belongs to the type-I magnetic point group, which is identical to the crystallographic point group $C_{2h}$ that contains one two-fold rotation axis ($C_{2}$) and one mirror plane ($M$) [see Fig.~\ref{fig1}(h)].  As mentioned before, the $M$ symmetry allows only for those components of the IAHC that are perpendicular to the mirror plane (i.e., along the current $C_{2}$ axis), therefore, $\sigma^{z}$ should be always zero but $\sigma^{x}$ and $\sigma^{y}$ are generally finite for $\theta= 0^{\circ}$ (for  current Cartesian coordinates).  If $\theta= \frac{2\pi}{3}$ or $\frac{5\pi}{3}$, the mirror plane is parallel to the $yz$ plane and renders only $\sigma^{x}$ potentially nonzero.  Similarly, $\bar{1}$ is also a type-I magnetic point group that is identical to the crystallographic group $C_{i}$.  Since all components $\Omega_{n}^{x}$, $\Omega_{n}^{y}$, and $\Omega_{n}^{z}$ are even with respect to the spatial inversion symmetry $I$, the group $C_{i}$ imposes no restrictions on the shape of $\boldsymbol{\sigma}$, allowing all components to be finite.  Finally, $2^{\prime}/m^{\prime}=C_{i} \oplus T(C_{2h}-C_{i})$ is a type-III magnetic point group containing one $TM$ and one $TC_{2}$ operation [see Figs.~\ref{fig1}(i) and~\ref{fig1}(j)].  There are two scenarios: if $\theta= \frac{\pi}{6}$ [Fig.~\ref{fig1}(i)], $TM$ (or $TC_{2}$) symmetry forces $\sigma^{x}$ to vanish but facilitates nonzero $\sigma^{y}$ and $\sigma^{z}$; if $\theta= \frac{\pi}{2}$ [Fig.~\ref{fig1}(j)], the principal axis of both symmetry operations changes ($M$ is neither parallel to $yz$ nor $zx$ plane) such that all entries $\sigma^{x}$, $\sigma^{y}$ and $\sigma^{z}$ are finite.  The other cases of $\theta= \frac{7\pi}{6}$ and $\frac{5\pi}{6}$ are identical to $\theta= \frac{\pi}{6}$ and $\frac{\pi}{2}$, respectively.  In summary, all tensor components of $\boldsymbol{\sigma}$ are allowed (except for some particular $\theta$) for the left-handed spin chirality owing to the reduced symmetry as compared to the systems with right-handed spin chirality.
	
	In the above discussion, all zero and potentially nonzero elements of the IAHC tensor are identified based on the underlying magnetic point groups. Alternatively, these results can also be obtained by following the Neumann principle, i.e., by applying all symmetry operations of the corresponding point group to the conductivity tensor~\cite{Seemann2015}. This method has been implemented in a computer program~\cite{Zelezny2017a,Zelezny2018a}, which generates the shape of linear response tensors (IAHC or intrinsic spin Hall conductivity) in a given coordinate system. Another useful analysis tool is the so-called cluster multipole theory~\cite{Suzuki2017,Suzuki2018}, which is capable of uncovering the hidden AHE in AFMs by evaluating the cluster multipole moment that behaves as a macroscopic magnetic order.  For instance, although the cluster dipole moments (i.e., the net magnetization from the conventional understanding) vanish in noncollinear AFMs (e.g., Mn$_{3}X$ and Mn$_{3}Y$), the emerging cluster octupole moments lead to a finite AHE.
		
	
	\begin{figure*}
		\includegraphics[width=2\columnwidth]{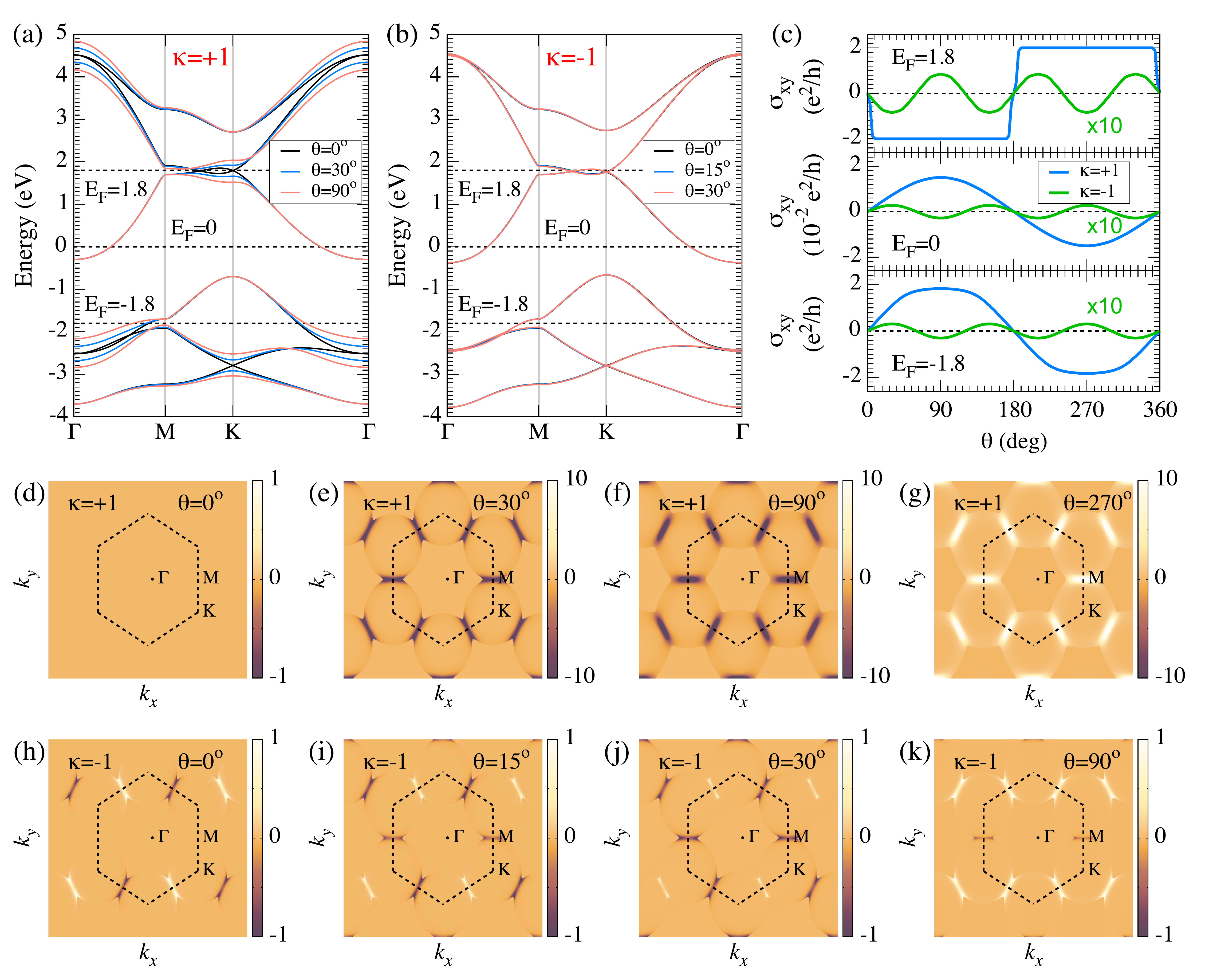}
		\caption{(Color online) (a)~Band structures of the kagome lattice with the spin orders $\kappa=+1$ and $\theta=0^{\circ}$, $30^{\circ}$, $90^{\circ}$. (b)~Band structures of the kagome lattice with the spin orders $\kappa=-1$ and $\theta=0^{\circ}$, $15^{\circ}$, $30^{\circ}$. (c)~IAHC of the kagome lattice as a function of $\theta$ for $\kappa=\pm1$ states for the three positions of the Fermi energy $E_{F}$ at 1.8~eV (top panel), 0~eV (middle panel), and $-$1.8~eV (bottom panel).  The curves of the $\kappa=-1$ state (green lines) are scaled by a factor of 10.  (d)-(g)~Berry curvature $\Omega_{xy}(\bm{k})$ with $\kappa=+1$ and $\theta=0^{\circ}$, $30^{\circ}$, $90^{\circ}$, $270^{\circ}$ at $E_{F}=-1.8$~eV.  (h)-(k)~Berry curvature $\Omega_{xy}(\bm{k})$ with $\kappa=-1$ and $\theta=0^{\circ}$, $15^{\circ}$, $30^{\circ}$, $90^{\circ}$ at $E_{F}=-1.8$~eV. Dotted lines in panels (d)-(k) indicate the first Brillouin zone.}
		\label{fig2}
	\end{figure*}
	
	\section{Tight-binding model}\label{sec3}
	
	Group theory is particularly powerful to identify the tensor shape of the IAHC, but it provides no insights into the magnitude of the allowed elements, which will depend strongly on details of the electronic structure. In this light, tight-binding models and first-principles calculations are valuable tools to arrive at quantitative predictions.  In this section, we consider a double-exchange $s$-$d$ model that describes itinerant $s$ electrons interacting with local $d$ magnetic moments on the kagome lattice, which refers to the (111) plane of cubic Mn$_{3}X$N.  Following Ref.~\onlinecite{H-Chen2014}, the Hamiltonian is written as
	\begin{eqnarray}\label{eq:Hamiltonian}
	H & = & t\sum_{\left<ij\right>\alpha}c_{i\alpha}^{\dagger}c_{j\alpha}-J\sum_{i\alpha\beta}\left(\boldsymbol{\tau}_{\alpha\beta}\cdot\boldsymbol{S}_{i}\right)c_{i\alpha}^{\dagger}c_{i\beta} \nonumber \\
	&  & + it_{\text{SO}}\sum_{\left<ij\right>\alpha\beta}\nu_{ij}\left(\boldsymbol{\tau}_{\alpha\beta}\cdot\boldsymbol{n}_{ij}\right)c_{i\alpha}^{\dagger}c_{i\beta},
	\end{eqnarray}
	where $c_{i\alpha}^{\dagger}$ ($c_{i\alpha}$) is the electron creation (annihilation) operator on site $i$ with spin $\alpha$, and $\boldsymbol{\tau}$ is the vector of Pauli matrices, and $\left\langle ij\right\rangle$ restricts the summation to nearest-neighbor sites.  The first term is the nearest-neighbor hopping with the transfer integral $t$.  The second term is the on-site exchange coupling between the conduction electron and the localized spin moment $\boldsymbol{S}_{i}$, and $J$ is the Hund's coupling strength.  The third term is the SOC effect with coupling strength $t_{\text{SO}}$, $\nu_{ij}$ is the antisymmetric 2D Levi-Civita symbol (with $\nu_{12}=\nu_{23}=\nu_{31}=1$), and $\boldsymbol{n}_{ij}$ is an in-plane vector perpendicular to the line from site $j$ to site $i$~\cite{H-Chen2014}.  In the following calculations, we set $J=1.7t$ and $t_{\text{SO}}=0.2t$.
	
	We first discuss the band structure, the IAHC, and the Berry curvature of the system with right-handed spin chirality ($\kappa=+1$), plotted in Figs.~\ref{fig2}(a),~\ref{fig2}(c), and~\ref{fig2}(d-g), respectively. The band structure significantly changes from $\theta=0^{\circ}$ (R1 phase, $\bar{3}1m$), to $30^{\circ}$ (R2 phase, $\bar{3}$), and to $90^{\circ}$ (R3 phase, $\bar{3}1m^{\prime}$).  If $\theta=0^{\circ}$, two band crossings around 1.8 eV appear at the $K$ point and along the $M$-$K$ path, respectively.  This band structure is identical to the one without SOC~\cite{H-Chen2014}, because the SOC term in Eq.~\eqref{eq:Hamiltonian} plays no effect in the spin configuration of $\theta=0^{\circ}$ in the sense that the left-handed and right-handed environments of an electron hopping between nearest neighbors are uniform.  Accordingly, the Berry curvature $\Omega_{xy}(\bm{k})$ vanishes everywhere in the Brillouin zone [Fig.~\ref{fig2}(d)].  The band degeneracy is split when $\theta\neq0^{\circ}$ and with increasing $\theta$, the band gap at the $K$ point enlarges significantly, while the one at the $M$ point shrinks slightly.  In order to disentangle the dependence of the IAHC on the band structure, the IAHC is calculated at different Fermi energies ($E_{F}$) including 1.8~eV, 0~eV, and $-$1.8~eV, shown in Fig.~\ref{fig2}(c).  In all three cases, the IAHC exhibits a period of $2\pi$ in $\theta$, and the values for $E_{F}=\pm1.8$~eV are two order of magnitude larger than the ones at $E_{F}=0$~eV.  The large IAHC originates from the small band gap at the $M$ point since the Berry curvature shows sharp peaks there [see Figs.~\ref{fig2}(e-g)].  For $E_{F}=-1.8$~eV and $0$~eV, the largest IAHC occurs for $\theta=90^{\circ}$ and $270^{\circ}$. The case of $E_{F}=1.8$ eV is special since the IAHC is quantized to $\pm2e^{2}/\hbar$ in a broad range of $\theta$, revealing the presence of quantum anomalous Hall state in coplanar noncollinear AFMs.
	
	For the left-handed spin chirality ($\kappa=-1$), the band structure, the IAHC, and the Berry curvature are plotted in Figs.~\ref{fig2}(b),~\ref{fig2}(c), and~\ref{fig2}(h-k), respectively.  The band structure hardly changes from $\theta=0^{\circ}$ ($2/m$), to $15^{\circ}$ ($\bar{1}$), and to $30^{\circ}$ ($2^{\prime}/m^{\prime}$).  If $\theta=0^{\circ}$, the Berry curvature $\Omega_{xy}(\bm{k})$ is odd for the group $2/m$ [Fig.~\ref{fig2}(h)] such that the IAHC $\sigma_{xy}$ is zero when integrating the $\Omega_{xy}(\bm{k})$ over the entire Brillouin zone.  With increasing $\theta$, the IAHC reaches its maximum at $\theta=30^{\circ}$ and exhibits a period of $\frac{2\pi}{3}$ [Fig.~\ref{fig2}(c)].  Similarly to the $\kappa=+1$ state, the IAHC at $E_{F}=\pm1.8$~eV is two orders of magnitude larger than at $E_{F}=0$~eV.  However, the IAHC of $\kappa=-1$ state is much smaller than that of $\kappa=+1$ state [Fig.~\ref{fig2}(c)].  This is understood based on the Berry curvature shown in Figs.~\ref{fig2}(i-k), which reveals that $\Omega_{xy}(\bm{k})$ at the three $M$ points has different signs (two negative and one positive, or two positive and one negative) due to the reduced symmetry in the $\kappa=-1$ state, in contrast to the same sign in the $\kappa=+1$ state [Figs.~\ref{fig2}(e-g)].
	
	The tight-binding model used here is constructed on a two-dimensional kagome lattice, for which the  $\sigma_{yz}$ and $\sigma_{zx}$ components vanish. Although the model is rather simple, the following qualitative results are useful: (1) the IAHC turns out to be large if the Fermi energy lies in a small band gap as outlined in previous theoretical work~\cite{YG-Yao2004}; (2) $\sigma_{xy}$ has a period of $2\pi$ ($\frac{2\pi}{3}$) in $\theta$ for right-handed (left-handed) spin chirality; (3) For structures with right-handed spin chirality, $\sigma_{xy}$ is much larger than for the left-handed case.
	
	\section{First-principles calculations}\label{sec4}
	
	In this section, by computing explicitly the electronic structure of the Mn$_3X$N compounds with different spin orders, we first demonstrate that key properties of these systems follow the qualitative conclusions drawn from the discussed tight-binding model.  Then, we present the values of the computed magnetic anisotropy energy (MAE) and the IAHC of the Mn$_{3}X$N compounds.  The obtained in-plane easy spin orientations are comparable to the previous reports~\cite{Mochizuki2018}, while the IAHC is found to depend strongly on the spin order, in agreement with the above tight-binding results.  Taking Mn$_{3}$NiN as an example, we further discuss the longitudinal and transverse optical conductivity, which are key to evaluating the MOE.  Finally, the spin-order dependent MOKE and MOFE as well as their anisotropy are explored. Computational details of the first-principles calculations are given in Appendix~\ref{appendix}.
	
	\begin{table}[b!]
		\caption{Magnetic anisotropy constant ($K_\text{eff}$) and the maximum of IAHC for Mn$_{3}X$N ($X$ = Ga, Zn, Ag, and Ni).  The IAHC is listed in the order of $\sigma_{yz}$, $\sigma_{zx}$, and $\sigma_{xy}$.  For the $\kappa=+1$ state, $\sigma_{xy}$ reaches its maximum at $\theta=90^{\circ}$.  For the $\kappa=-1$ state, $\sigma_{yz}$, $\sigma_{zx}$, and $\sigma_{xy}$ reach their maxima at $\theta=120^{\circ}$, $30^{\circ}$, and $30^{\circ}$, respectively.}
		\label{tab2}
		\begin{ruledtabular}
			\begingroup
			\setlength{\tabcolsep}{4.5pt} 
			\renewcommand{\arraystretch}{1.5} 
			\begin{tabular}{ccccc}
				
				\multicolumn{1}{c}{} &
				\multicolumn{2}{c}{$K_\text{eff}$ (meV/cell)} & 
				\multicolumn{2}{c}{IAHC (S/cm)}  \\
				\cline{2-3}
				\cline{4-5}
				
				System & $\kappa=+1$ & $\kappa=-1$ & $\kappa=+1$ & $\kappa=-1$ \\
				
				\hline
				
				Mn$_{3}$GaN  & 0.52 & 0.26 & 0, 0, $-$99 & 59, $-$67, $-$5 \\
				
				Mn$_{3}$ZnN  & 0.43 & 0.21 & 0, 0, $-$232 & 156, $-$174, 23 \\
				
				Mn$_{3}$AgN  & 0.15 & 0.08 & 0, 0, $-$359 & 344, $-$314, 72 \\
				
				Mn$_{3}$NiN  & $-$0.18 & $-$0.09 & 0, 0, $-$301 & 149, $-$134, 5 \\
				
			\end{tabular}
			\endgroup
		\end{ruledtabular}
	\end{table}
	
	\begin{figure}
		\includegraphics[width=\columnwidth]{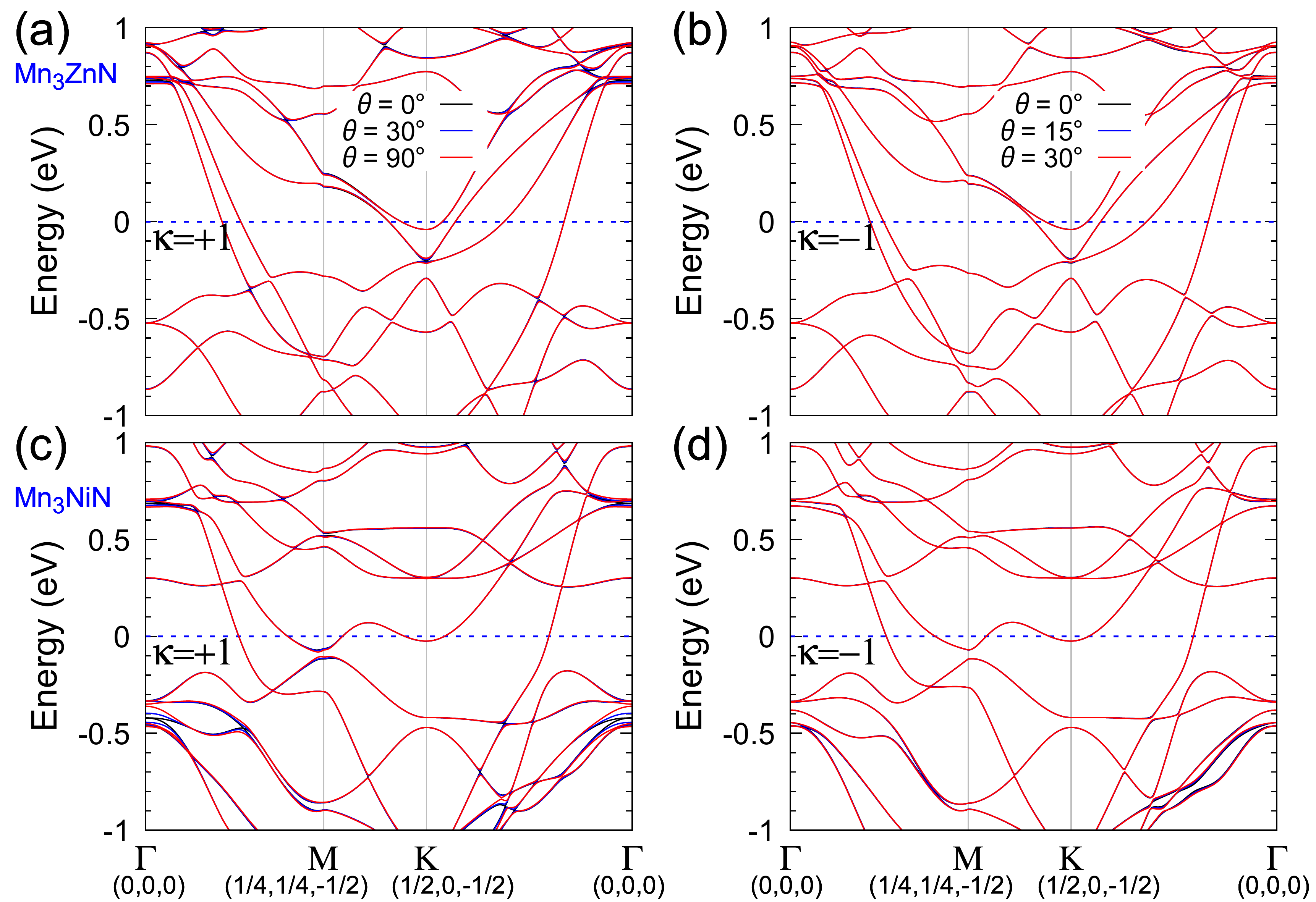}
		\caption{(Color online) The first-principles band structures of (a,b)~Mn$_{3}$ZnN and (c,d)~Mn$_{3}$NiN for different spin orders ($\theta=0^{\circ}$, $30^{\circ}$, and $90^{\circ}$ for the right-handed state with $\kappa=+1$, and $\theta=0^{\circ}$, $15^{\circ}$, and $30^{\circ}$ for the left-handed state of opposite spin chirality). The $k$-path lies within the (111) plane.}
		\label{fig3}
	\end{figure}

	\begin{figure}
		\includegraphics[width=\columnwidth]{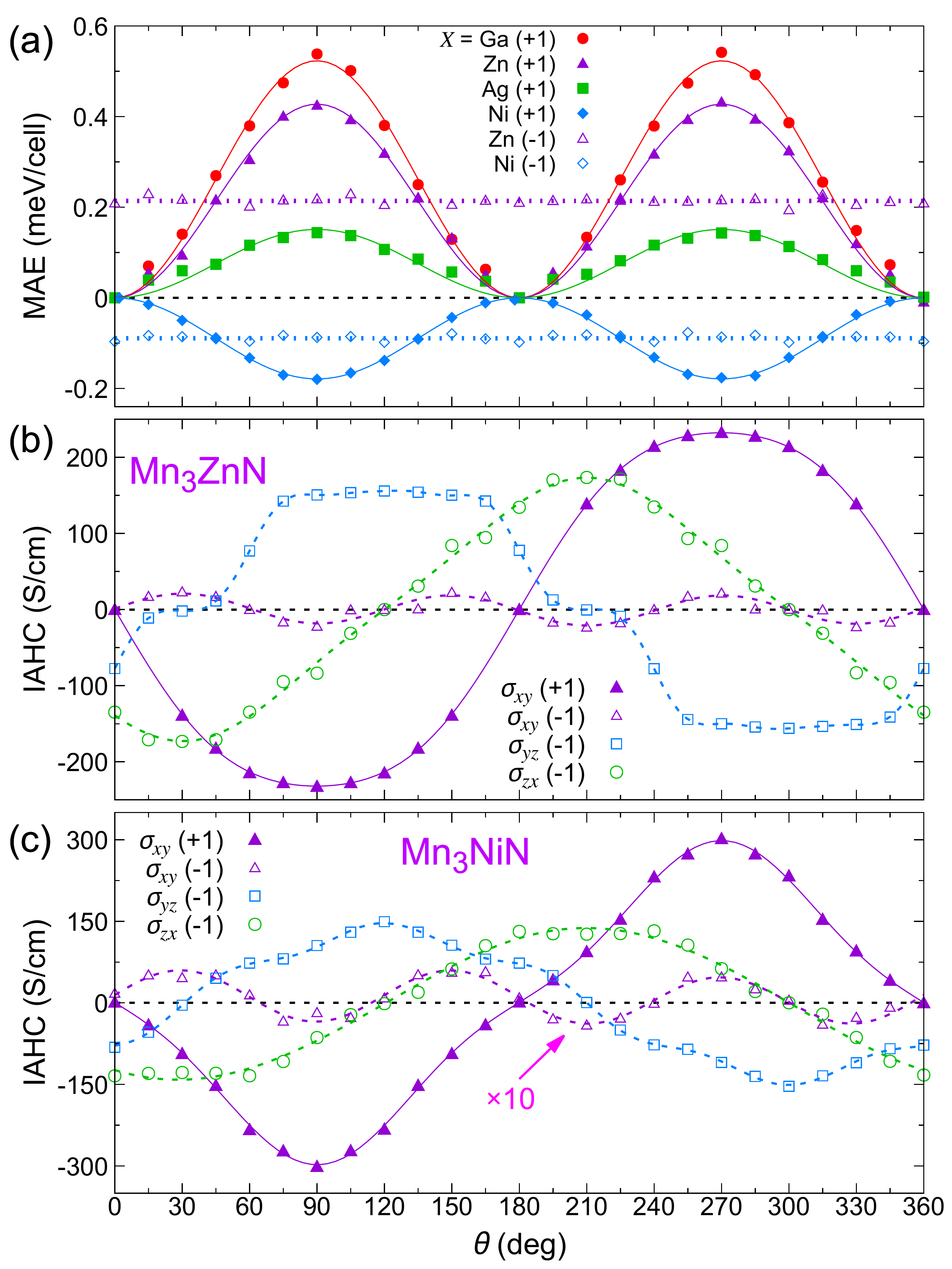}
		\caption{(Color online) (a) Magnetic anisotropy energy of Mn$_{3}X$N ($X$ = Ga, Zn, Ag, and Ni) as a function of the azimuthal angle $\theta$.  The results for left-handed spin chirality ($\kappa=-1$) are only shown in Mn$_{3}$ZnN and Mn$_{3}$NiN as the representatives.  The solid and dotted lines are expressed by $\text{MAE}(\theta)=K_{\text{eff}}\sin^{2}(\theta)$ and $\text{MAE}=K_{\text{eff}}/2$, respectively.  (b), (c) The IAHC of Mn$_{3}$ZnN  and Mn$_{3}$NiN as a function of the azimuthal angle $\theta$ for both right-handed ($\kappa=+1$) and left-handed ($\kappa=-1$) spin chiralities.  The solid and dotted lines are the polynomial fits to the data.}
		\label{fig4}
	\end{figure}
	
	\subsection{Electronic structure}\label{sec4-0}
	Figure~\ref{fig3} illustrates the first-principles band structures of the Mn$_{3}X$N systems, taking Mn$_{3}$ZnN and Mn$_{3}$NiN as two prototypical examples. While the electronic structure of the left-handed state with $\kappa=-1$ hardly changes as the spin-rotation angle $\theta$ is tuned, the right-handed state of opposite vector spin chirality is rather sensitive to details of the noncollinear spin configuration. Specifically, the calculated electronic structure for the $\kappa=+1$ state reveals that the band degeneracy (e.g., at the $\Gamma$ point) is lifted for $\theta\neq0^{\circ}$, and the magnitude of the band splitting increases with the spin-rotation angle. These features are in a very good qualitative agreement with the tight-binding results [see Figs.~\ref{fig2}(a) and~\ref{fig2}(b)], which roots in the fact that the (111) planes of Mn$_{3}X$N compounds and the 2D Kagome lattice considered in the previous sections have common symmetries.

	\subsection{Intrinsic anomalous Hall conductivity and magnetic anisotropy energy}\label{sec4-1}
	
	The MAE is one of the most important parameters that characterizes a magnetic material.  In FMs, the MAE refers to the total energy difference between easy- and hard-axis magnetization directions. In the noncollinear AFMs that we consider here, we define the MAE as the total energy difference between different spin orders, given by
	\begin{equation}\label{eq:MAE}
	\text{MAE}(\theta)=E_{\kappa=\pm1,\theta\neq0^{\circ}}-E_{\kappa=+1,\theta=0^{\circ}},
	\end{equation}
	where the spin order with $\kappa=+1$ and $\theta=0^{\circ}$ is set as the reference state.  The calculated MAE of Mn$_{3}X$N is plotted in Fig.~\ref{fig4}(a). For the $\kappa=+1$ state, the MAE can be fitted well to the uniaxial anisotropy $K_{\text{eff}}\sin^{2}(\theta)$, where $K_{\text{eff}}$ is the magnetic anisotropy constant listed in Tab.~\ref{tab2}. Compared to traditional Mn-based alloys, the value of $K_{\text{eff}}$ in Mn$_{3}X$N is comparable in magnitude MnPt (0.51 meV/cell)~\cite{Umetsu2006}, MnPd ($-$0.57 meV/cell)~\cite{Umetsu2006}, MnNi ($-$0.29 meV/cell)~\cite{Umetsu2006}, and MnRh ($-$0.63 meV/cell)~\cite{Umetsu2006}, but are one order of magnitude smaller than in MnIr ($-$7.05 meV/cell)~\cite{Umetsu2006}, Mn$_3$Pt (2.8 meV/cell)~\cite{Kota2008}, and Mn$_3$Ir (10.42 meV/cell)~\cite{Szunyogh2009}.  For the $\kappa=-1$ state, the MAE is approximately constant with a value of $K_{\text{eff}}/2$, indicating the vanishing in-plane anisotropy energy that leads to a relatively easy rotation of the spins within the (111) plane.  This feature has also been found in other noncollinear AFMs such as Mn$_{3}$Ir~\cite{Szunyogh2009}, Mn$_{3}$Ge~\cite{Nagamiya1982}, and Mn$_{3}$Sn~\cite{Nagamiya1982,Tomiyoshi1982,Nakatsuji2015}.
	
	\begin{figure*}
		\includegraphics[width=0.95\textwidth]{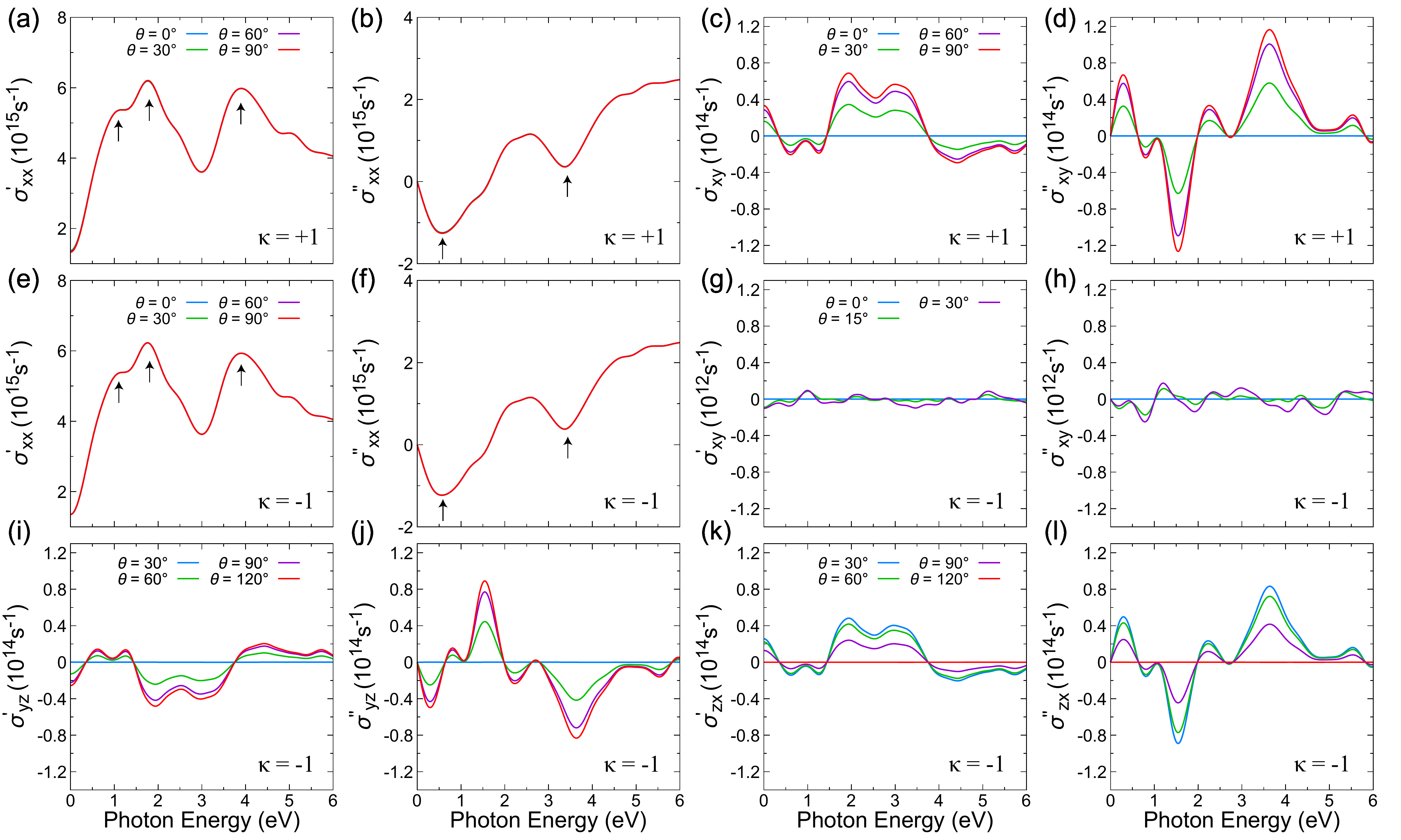}
		\caption{(Color online) Energy dependence of the optical conductivity in Mn$_{3}$NiN.  (a-b)~Real and imaginary parts of $\sigma_{xx}$ for the $\kappa=+1$ state. Characteristic peaks and valleys are marked by black arrows.  (c-d)~Real and imaginary parts of $\sigma_{xy}$ for the $\kappa=+1$ state. (e-f)~The real and imaginary parts of $\sigma_{xx}$ for the $\kappa=-1$ state.  (g-h), (i-j), (k-l)~The real and imaginary parts of $\sigma_{xy}$, $\sigma_{yz}$, and $\sigma_{zx}$ for the $\kappa=-1$ state.}
		\label{fig5}
	\end{figure*}
	
	Fig.~\ref{fig4}(a) reveals that $\text{MAE}(\theta)=\text{MAE}(\theta+\pi)$, implying that the ground state of $120^{\circ}$ triangular spin order has a discrete two-fold degeneracy~\cite{H-Chen2014}.  For the $\kappa=+1$ state, Mn$_{3}$GaN and Mn$_{3}$ZnN obviously prefer the R1 phase ($\theta=0^{\circ}$ or $180^{\circ}$), which is in full accordance with the $\Gamma^{5g}$ spin configuration identified in Ref.~\onlinecite{Mochizuki2018} using a classical spin model with frustrated exchange interactions and magnetic anisotropy.  As the spin configuration is closely related to the number of valence electrons $n_{\nu}$ in the $X$ ion, Mochizuki \textit{et al.}~\cite{Mochizuki2018} propose a mixture of the $\Gamma^{5g}$ and the $\Gamma^{4g}$ spin patterns in Mn$_{3}$AgN and Mn$_{3}$NiN due to the smaller $n_{\nu}$ (weaker $X$-ion crystal field) as compared to that of Mn$_{3}$GaN and Mn$_{3}$ZnN.  In the present calculations, Mn$_{3}$AgN still hosts the $\Gamma^{5g}$ spin configuration but has a much smaller MAE compared to Mn$_{3}$GaN and Mn$_{3}$ZnN, while Mn$_{3}$NiN favors the $\Gamma^{4g}$ spin configuration (R3 phase, $\theta=90^{\circ}$ or $270^{\circ}$).  Our calculated MAE is a monotonic function of $n_{\nu}$, i.e., Ni $(n_{\nu}=0)<$ Ag $(n_{\nu}=1)<$ Zn $(n_{\nu}=2)<$ Ga $(n_{\nu}=3)$, which provides a clear interpretation for the systematic evolution of the magnetic orders in Mn$_{3}X$N.  On the other hand, the $\kappa=-1$ state of Mn$_{3}X$N has not been considered in previous works, while we find that it could exist for particular values of $\theta$.  For example, the $\kappa=-1$ state in Mn$_{3}$NiN has the favorable energy in three segments of $\theta$: $[0^{\circ},45^{\circ})$, $(135^{\circ},225^{\circ})$, and $(315^{\circ},360^{\circ}]$.  In the light of recent experiments on Mn$_{3}$Sn~\cite{Nakatsuji2015,Ikhlas2017} and Mn$_{3}$Ge~\cite{Nayak2016,Kiyohara2016}, an external magnetic field may be used to tune the spin orientation by coupling to the weak in-plane magnetic moment. This finding  enriches the spectrum of possible magnetic orders in Mn$_{3}X$N compounds.
	
	The IAHC of Mn$_{3}$ZnN and Mn$_{3}$NiN with different spin orders is illustrated in Figs.~\ref{fig4}(b) and~\ref{fig4}(c), respectively.  The component $\sigma_{xy}$ displays a period of $2\pi$ ($\frac{2\pi}{3}$) in $\theta$ for the $\kappa=+1$ ($\kappa=-1$) state, and its magnitude in the $\kappa=+1$ state is much larger than that of the $\kappa=-1$ state, in excellent agreement with the tight-binding results. From the group theoretical analysis  we showed that $\sigma_{yz}$ and $\sigma_{zx}$ are allowed in $\kappa=-1$ state, which is confirmed by our first-principles results. Moreover, we observe that both $\sigma_{yz}$ and $\sigma_{zx}$ display a period of $2\pi$ in $\theta$ and their magnitudes are much larger than that of $\sigma_{xy}$.  The maximum of IAHC for $\kappa=\pm1$ states is summarized in Tab.~\ref{tab2}. Overall, the obtained magnitude of the IAHC in the studied family of compounds is comparable or even larger than that in other noncollinear AFMs like Mn$_3X$~\cite{H-Chen2014,Y-Zhang2017} and Mn$_{3}Y$~\cite{Kubler2014,GY-Guo2017,Y-Zhang2017,Nakatsuji2015,Nayak2016,Kiyohara2016,Ikhlas2017}.  In contrast to the MAE, the IAHC follows the relation  $\boldsymbol{\sigma}(\theta)=-\boldsymbol{\sigma}(\theta+\pi)$, which manifests that the spin state at $\theta+\pi$ is the time-reversed counterpart of the order at $\theta$ and the IAHC is odd under time-reversal symmetry.

	\begin{figure*}
		\includegraphics[width=\textwidth]{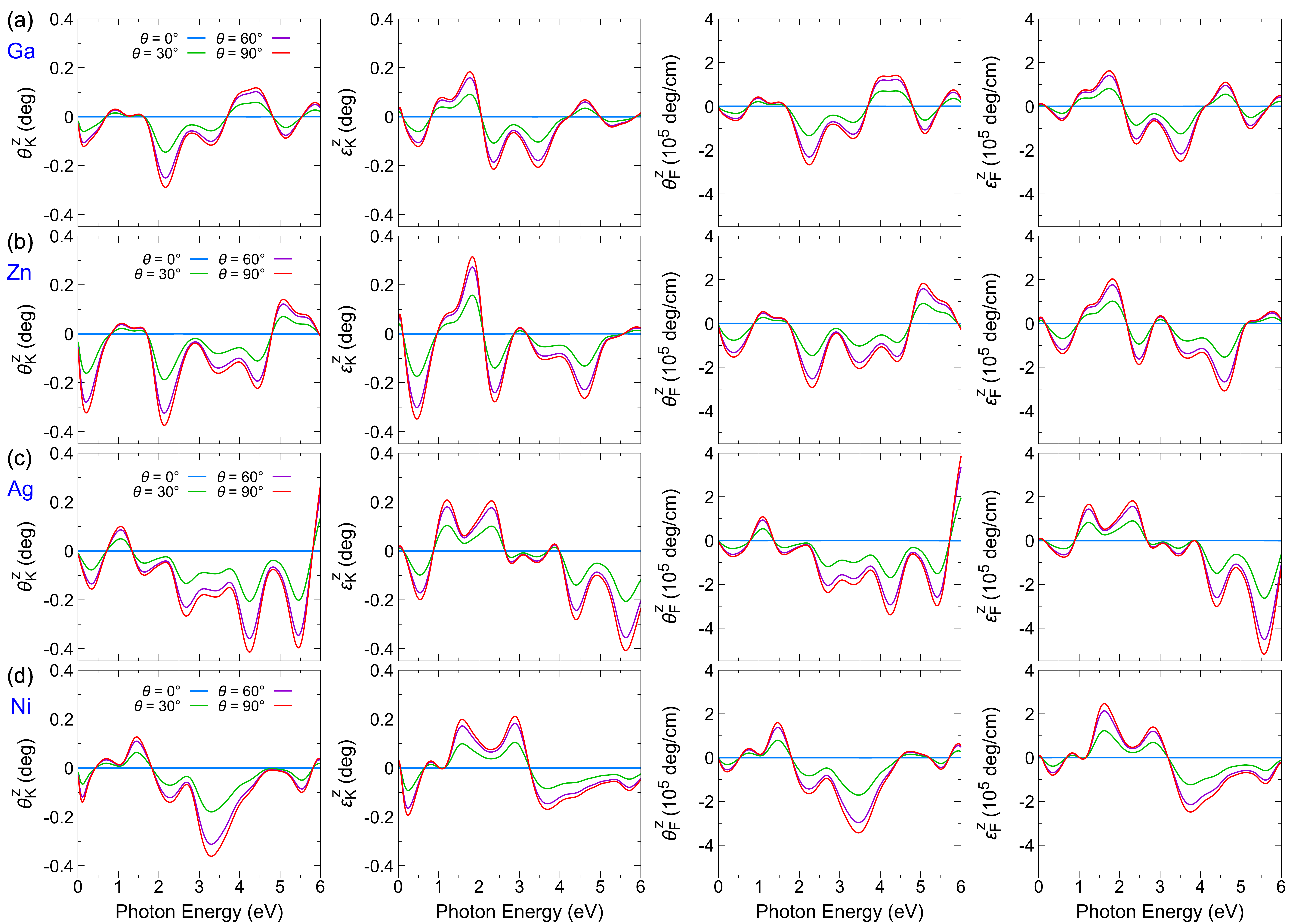}
		\caption{(Color online) Magneto-optical spectra of Mn$_{3}X$N for $X=$ Ga~(a), Zn~(b), Ag~(c), and Ni~(d) in the $\kappa=+1$ spin configuration.  The panels from left to right show Kerr rotation angle $\theta^{z}_{K}$, Kerr ellipticity $\varepsilon^{z}_{K}$, Faraday rotation angle $\theta^{z}_{F}$, and Faraday ellipticity $\varepsilon^{z}_{F}$, respectively.}
		\label{fig6}
	\end{figure*}

	\subsection{Optical conductivity}\label{sec4-2}
	
	Before proceeding to the MOE, we evaluate the optical conductivity as a key quantity that comprises the MOE. Expanding on the expressions for the IAHC [Eqs.~\eqref{eq:IAHC} and~\eqref{eq:BerryCur}], the optical conductivity can be written down as
	\begin{eqnarray}\label{eq:optical}
	\sigma_{\alpha\beta}(\omega) & = & \sigma^{\prime}_{\alpha\beta}(\omega) + i \sigma^{\prime\prime}_{\alpha\beta}(\omega) \nonumber \\
	& = & \hbar e^{2}\int\frac{d^{3}k}{(2\pi)^{3}}\sum_{n\neq n^{\prime}}\left[f_{n}(\bm{k})-f_{n^{\prime}}(\bm{k})\right] \nonumber \\
	&  & \times\frac{\textrm{Im}\left[\left\langle \psi_{n\bm{k}}|v_{\alpha}|\psi_{n^{\prime}\bm{k}}\right\rangle \left\langle \psi_{n^{\prime}\bm{k}}|v_{\beta}|\psi_{n\bm{k}}\right\rangle\right] }{(\hbar\omega_{n\bm{k}}-\hbar\omega_{n^{\prime}\bm{k}})^{2}-(\hbar\omega+i\eta)^{2}},
	\end{eqnarray}
	where the superscript $^{\prime}$ ($^{\prime\prime}$) of $\sigma_{\alpha\beta}$ denotes its the real (imaginary) part, $\eta$ is an adjustable smearing parameter in units of energy, and $\hbar\omega$ is the photon energy.  Due to the found similarity in the results among all studied systems, we take  Mn$_{3}$NiN as a representative example for discussing the optical conductivity (Fig.~\ref{fig5}).  The real part of the diagonal element, $\sigma^{\prime}_{xx}$ [see Figs.~\ref{fig5}(a) and.~\ref{fig5}(e)], measures the average in the absorption of left- and right-circularly polarized light. The spectrum exhibits one absorptive peak at 1.8~eV with a shoulder at 1.1~eV and another absorptive peak at 3.9~eV.   The imaginary part of the diagonal element, $\sigma^{\prime\prime}_{xx}$ [see Figs.~\ref{fig5}(b) and.~\ref{fig5}(f)], is the dispersive part of the optical conductivity, revealing two distinct valleys at 0.6 eV and 3.4 eV.  Obviously, $\sigma_{xx}$ is not affected by the spin order (neither spin chirality $\kappa$ nor azimuthal angle $\theta$).  A similar behavior has been found in Mn$_{3}X$~\cite{WX-Feng2015}, where $\sigma_{xx}$ is identical for T1 and T2 spin structures.  From the symmetry analysis~\cite{Seemann2015}, it should hold that $\sigma_{xx}=\sigma_{yy}\neq\sigma_{zz}$ for the magnetic point groups  $\bar{3}1m$, $\bar{3}$, and $\bar{3}1m^{\prime}$ in the $\kappa=+1$ state, whereas $\sigma_{xx}\neq\sigma_{yy}\neq\sigma_{zz}$ for the magnetic point groups of $2/m$, $\bar{1}$, and $2^{\prime}/m^{\prime}$ in the $\kappa=-1$ state.  However, all diagonal elements are approximately equal in our calculations, i.e., we observe that $\sigma_{xx}\approx\sigma_{yy}\approx\sigma_{zz}$. This promotes the optical isotropy in the Mn$_{3}X$N family.
	
	In contrast to the diagonal entries, the off-diagonal elements displayed in Figs.~\ref{fig5}(c,d) and~\ref{fig5}(g--l) depend significantly on the spin order.  For the $\kappa=+1$ state [Figs.~\ref{fig5}(c,d)], $\sigma_{xy}(\omega)$ vanishes if $\theta=0^{\circ}$, but it increases with the increasing $\theta$ and reaches its maximum at $\theta=90^{\circ}$.  For the $\kappa=-1$ state [Figs.~\ref{fig5}(g--l)], all three off-diagonal elements $-$ $\sigma_{xy}(\omega)$, $\sigma_{yz}(\omega)$, and $\sigma_{zx}(\omega)$ $-$ can be nonzero and they peak at $\theta=30^{\circ}$, $120^{\circ}$, and $30^{\circ}$, respectively. Furthermore, $\sigma_{xy}(\omega)$ is at least two orders of magnitude smaller than $\sigma_{yz}(\omega)$ and $\sigma_{zx}(\omega)$.  The overall trend of $\sigma_{xy}(\omega)$ depending on the spin order is very similar to that of the IAHC in Fig.~\ref{fig4}(c).

	\begin{figure*}
		\includegraphics[width=\textwidth]{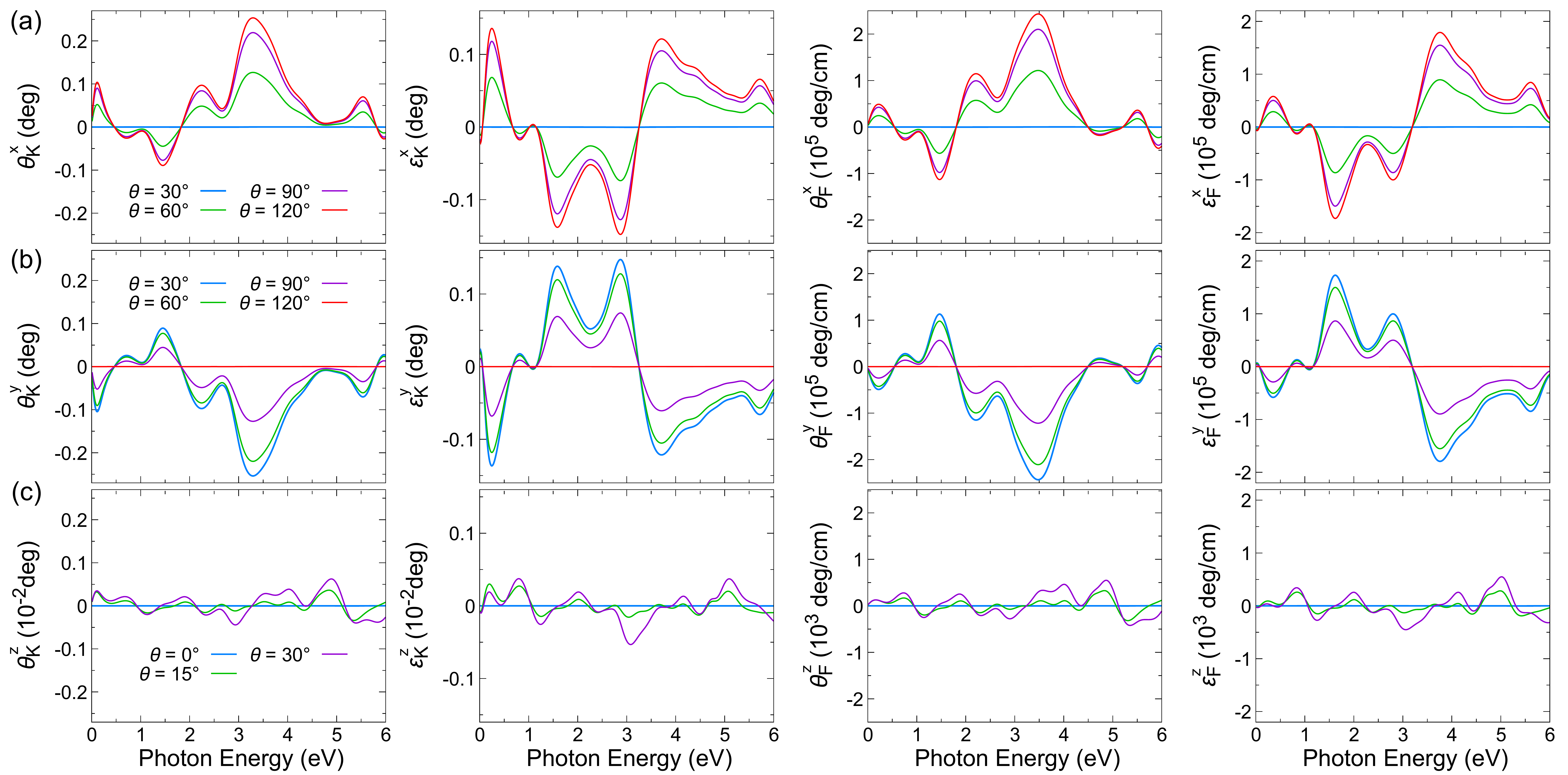}
		\caption{(Color online) The magneto-optical spectra of Mn$_{3}$NiN for $\kappa=-1$ state: $\phi^{x}_{K,F}$ (a), $\phi^{y}_{K,F}$ (b), and $\phi^{z}_{K,F}$ (c).  The panels from left to right are Kerr rotation angle, Kerr ellipticity, Faraday rotation angle, and Faraday ellipticity, respectively.}
		\label{fig7}
	\end{figure*}

	\subsection{Magneto-optical Kerr and Faraday effects}\label{sec4-3}
	
	We now turn to the magneto-optical Kerr and Faraday effects (MOKE and MOFE).  The characteristic Kerr rotation angle $\theta_{K}$ and the ellipticity $\varepsilon_{K}$ are typically combined into the complex Kerr angle given by~\cite{Kahn1969,GY-Guo1994,GY-Guo1995}
	\begin{eqnarray}\label{eq:kerr}
	\phi^{\gamma}_{K} & = & \theta^{\gamma}_{K}+i\varepsilon^{\gamma}_{K} \nonumber \\ 
	& = & \frac{-\nu_{\alpha\beta\gamma}\sigma_{\alpha\beta}}{\sigma_{0}\sqrt{1+i\left(4\pi/\omega\right)\sigma_{0}}},
	\end{eqnarray}
	where $\nu_{\alpha\beta\gamma}$ is the 3D Levi-Civita symbol with the Cartesian coordinates $\alpha,\beta,\gamma\in\{x,y,z\}$ and $\sigma_{0}=\frac{1}{2}(\sigma_{\alpha\alpha}+\sigma_{\beta\beta})\approx\sigma_{\alpha\alpha}$.  The complex Kerr angle expressed here, similarly to the IAHC, holds a pseudovector form, i.e., $\boldsymbol{\phi}_{K}=[\phi^{x}_{K},\phi^{y}_{K},\phi^{z}_{K}]$, which differentiates the Kerr angles when the incident light propagates along different crystallographic axes.  One can read from Eq.~\eqref{eq:kerr} that the longitudinal optical conductivity $\sigma_{\alpha\alpha}$ modulates the magnitude of the Kerr spectrum, while the transverse optical conductivity $\sigma_{\alpha\beta}$ determines key features of the Kerr spectrum.  For example, only the component $\phi^{z}_{K}$ is finite for the $\kappa=+1$ state, whereas all components $\phi^{x}_{K}$, $\phi^{y}_{K}$, and $\phi^{z}_{K}$ are nonzero in the $\kappa=-1$ configuration.  More importantly, $\phi^{x}_{K}\neq\phi^{y}_{K}\neq\phi^{z}_{K}$ implies the presence of magneto-optical anisotropy if the incident light propagates along $x$ ($01\bar{1}$), $y$ ($\bar{2}11$), and $z$ ($111$) axes (see Fig.~\ref{fig1}), respectively.  Similarly, the complex Faraday angle can be expressed as~\cite{Reim1990}
	\begin{eqnarray}\label{eq:faraday}
	\phi^{\gamma}_{F} & = & \theta^{\gamma}_{F}+i\varepsilon^{\gamma}_{F} \nonumber \\ 
	& = & \nu_{\alpha\beta\gamma}(n_{+}-n_{-})\frac{\omega l}{2c},
	\end{eqnarray}
	where $n_{\pm}=[1+\frac{4\pi i}{\omega}(\sigma_{\alpha\alpha}\pm i\sigma_{\alpha\beta})]^{1/2}$ are the complex refractive indices and $l$ is the thickness of the thin film.  Since $\sigma_{\alpha\alpha}$ is generally much larger than $\sigma_{\alpha\beta}$ (see Fig.~\ref{fig5}),  $n_{\pm}\approx[1+\frac{4\pi i}{\omega}\sigma_{\alpha\alpha}]^{1/2}\mp\frac{2\pi}{\omega}\sigma_{\alpha\beta}[1+\frac{4\pi i}{\omega}\sigma_{\alpha\alpha}]^{-1/2}$ (Ref.~\onlinecite{YM-Fang2018}) and consequently, the complex Faraday angle can be approximated as $\theta^{\gamma}_{F}+i\varepsilon^{\gamma}_{F} = -\nu_{\alpha\beta\gamma}\frac{2\pi l}{c}\sigma_{\alpha\beta}[1+\frac{4\pi i}{\omega}\sigma_{\alpha\alpha}]^{-1/2}$.  Therefore, the Faraday spectrum is also determined by $\sigma_{\alpha\beta}$.
	
	The magneto-optical Kerr and Faraday spectra for the spin order with $\kappa=+1$ in Mn$_{3}X$N are plotted in Fig.~\ref{fig6}, where only $\phi^{z}_{K,F}$ are shown since all other components vanish.  Taking Mn$_{3}$NiN as an example [Fig.~\ref{fig6}(d)], one can observe that the Kerr and Faraday spectra indeed inherit the behavior of the optical conductivity $\sigma_{xy}(\omega)$ [Figs.~\ref{fig5}(c) and~\ref{fig5}(d)].  For example, the Kerr and Faraday angles are zero when $\theta=0^{\circ}$, increase with increasing $\theta$, and reach their maximum at $\theta=90^{\circ}$.  This indicates that the symmetry requirements for MOKE and MOFE are the same as that for the optical Hall conductivity.  In addition, all Mn$_{3}X$N compounds considered here have similar Kerr and Faraday spectra, primarily due to their isostructural nature.  The Kerr rotation angles in Mn$_{3}X$N are comparable to the theoretical values in Mn$_{3}X$ ($0.2\sim0.6$ deg)~\cite{WX-Feng2015} and are larger than the experimental value in Mn$_{3}$Sn (0.02 deg)~\cite{Higo2018}.  The largest Kerr and Faraday rotation angles of respectively 0.42 deg and $4\times10^{5}$ deg/cm emerge in Mn$_{3}$AgN. This roots potentially in the stronger SOC of the Ag atom as compared to other lighter $X$ atoms.
	
	Figure~\ref{fig7} shows the magneto-optical Kerr and Faraday spectra for the $\kappa=-1$ state of Mn$_{3}$NiN.  Since all off-diagonal elements $\sigma_{yz}(\omega)$, $\sigma_{zx}(\omega)$, and $\sigma_{xy}(\omega)$ of the optical conductivity are nonzero for the $\kappa=-1$ state, the Kerr and Faraday effects will appear if the incident light propagates along any Cartesian axes.  This is in contrast to the case of the $\kappa=+1$ configuration, for which only the incident light along the $z$ axis generates finite $\phi^{z}_{K,F}$.  In Fig.~\ref{fig7}(a), $\phi^{x}_{K,F}$ are zero at $\theta=30^{\circ}$ but have the largest values at $\theta=120^{\circ}$, owing to the features in $\sigma_{yz}$ [Figs.~\ref{fig5}(i) and~\ref{fig5}(j)].  Moreover, the Kerr and Faraday rotation angles ($\theta^{x}_{K}$ and $\theta^{x}_{F}$) and the ellipticity ($\varepsilon^{x}_{K}$ and $\varepsilon^{x}_{F}$) resemble, respectively, the real part ($\sigma^{\prime}_{yz}$) and imaginary part ($\sigma^{\prime\prime}_{yz}$) of the corresponding off-diagonal conductivity element.  Compared to $\phi^{x}_{K,F}$, the angle $\phi^{y}_{K,F}$ in Fig.~\ref{fig7}(b) displays an opposite behavior in the sense that it has the largest values at $\theta=30^{\circ}$ but vanishes at $\theta=120^{\circ}$.  This is not surprising as the periods of $\sigma_{yz}$ and $\sigma_{zx}$ as a function of $\theta$ differ by $\frac{\pi}{2}$, which can be read from Fig.~\ref{fig4}(c) and Figs.~\ref{fig5}(i-l).  The angles $\phi^{z}_{K,F}$ shown in Fig.~\ref{fig7}(c) are two orders of magnitude smaller than $\phi^{x}_{K,F}$ and $\phi^{y}_{K,F}$, implying that very weak Kerr and Faraday effects are expected for the incident light along the $z$ axis.  From Figs.~\ref{fig6} and~\ref{fig7}, we conclude that the MOKE and MOFE depend strongly on the spin order, as in the case of the IAHC.

	\section{Summary}\label{sec5}
	
	In summary, using a group theoretical analysis, tight-binding modelling, and first-principles calculations, we have systematically investigated the spin-order dependent intrinsic anomalous Hall effect and magneto-optical Kerr and Faraday effects in Mn$_3X$N ($X$ = Ga, Zn, Ag, and Ni)  compounds, which are considered to be an important class of noncollinear antiferromagnets.  The symmetry-imposed shape of the anomalous Hall conductivity tensor is determined via the analysis of magnetic point groups, that is, only $\sigma_{xy}$ can be nonzero for the right-handed spin chirality ($\kappa=+1$) while finite $\sigma_{xy}$, $\sigma_{yz}$, and $\sigma_{zx}$ exist for the left-handed spin chirality ($\kappa=-1$).  Our tight-binding modelling confirms these results and further reveals that $\sigma_{xy}$ is a \textit{sine}-like function of the azimuthal angle $\theta$ with a period of $2\pi$ ($\frac{2\pi}{3}$) for the $\kappa=+1$ ($\kappa=-1$) state.  By examining the $\bm{k}$-resolved Berry curvature, we uncovered that the intrinsic anomalous Hall conductivity is generally large if the Fermi energy enters into the region with small band gaps formed at anticrossings.  The first-principles calculations reproduce all features of $\sigma_{xy}$ and further verify that $\sigma_{yz}$ and $\sigma_{zx}$ have a period of $2\pi$ for the $\kappa=-1$ state.  The intrinsic anomalous Hall conductivity shows a distinct relation of $\boldsymbol{\sigma}(\theta)=-\boldsymbol{\sigma}(\theta+\pi)$ due to its odd nature under time-reversal symmetry.  In addition, we have calculated the magnetic anisotropy energy which manifests as $K_{\textrm{eff}}$ $\sin^{2}$($\theta)$ for the $\kappa=+1$ state but remains nearly constant at $K_{\textrm{eff}}/2$ for the $\kappa=-1$ state.  A discrete two-fold energy degeneracy, i.e., $\text{MAE}(\theta)=\text{MAE}(\theta+\pi)$, is found in the noncollinear antiferromagnetic Mn$_3X$N.  Strikingly, our first-principles calculations reveal that the $\kappa=-1$ state could exist in Mn$_3X$N for certain values of $\theta$.
	
	The optical conductivities for $\kappa=\pm1$ states were explored, considering Mn$_3$NiN as a prototypical example.  We find that the spin order hardly affects the diagonal elements whereas it influences strongly the off-diagonal entries.  The optical isotropy is established since  $\sigma_{xx}(\omega)\approx\sigma_{yy}(\omega)\approx\sigma_{zz}(\omega)$, while magneto-optical anisotropy occurs inevitably as $\sigma_{xy}(\omega)\neq\sigma_{yz}(\omega)\neq\sigma_{zx}(\omega)$.  Finally, magneto-optical Kerr and Faraday effects are evaluated based on the optical conductivity.  The largest Kerr rotation angles in Mn$_3X$N amount to 0.4 deg, which is comparable to other noncollinear antiferromagnets, e.g., Mn$_{3}X$~\cite{WX-Feng2015} and Mn$_{3}$Sn~\cite{Higo2018}.  Since the optical Hall conductivity plays a major role for magneto-optical effects, the Kerr and Faraday spectra also display a spin-order dependent behavior.  Our work illustrates that complex noncollinear spin structures could be probed via anomalous Hall and magneto-optical effects measurements.

	\begin{acknowledgments}
		W. F. and Y. Y. acknowledge the support from the National Natural Science Foundation of China (Nos. 11874085 and 11734003) and the National Key R\&D Program of China (No. 2016YFA0300600).  W. F. also acknowledges the funding through an Alexander von Humboldt Fellowship.  Y.M., J.-P. H. and S.B. acknowledge  funding under SPP 2137 ``Skyrmionics" (project MO 1731/7-1), collaborative Research Center SFB 1238, and Y.M. acknowledges  funding  from project  MO  1731/5-1 of Deutsche Forschungsgemeinschaft (DFG).  G.-Y. G. is supported by the Ministry of Science and Technology and the Academia Sinica as well as NCTS and Kenda Foundation in Taiwan.  We acknowledge computing time on the supercomputers JUQUEEN and JURECA at J\"ulich Supercomputing Centre and JARA-HPC of RWTH Aachen University.
	\end{acknowledgments}

	\appendix
	\section{The details of first-principles calculations}\label{appendix}
	
	First-principles density functional theory calculations are performed using the Vienna \textit{ab initio} simulation package (\textsc{vasp})~\cite{Kresse1996a}, in which the accurate frozen-core full-potential projector augmented wave (PAW) method is used~\cite{Bloechl1994}.  The exchange-correlation potential is treated by the generalized-gradient approximation with the parameters of Perdew-Burke-Ernzerhof~\cite{Perdew1996}. The energy cut-off of 500 eV, the energy criterion of 10$^{-6}$ eV, and the $k$-mesh of 16$\times$16$\times$16 are used in the self-consistent calculations.  Spin-orbit interaction is explicitly included in fully relativistic projector augmented potentials.  A penalty functional is added to total energy in order to constrain the local spin moments along a desired direction.  The experimental lattice constants of 3.898 {\AA}, 3.890 {\AA}, 4.013 {\AA}, and 3.886 {\AA} are adopted for Mn$_3X$N ($X$ = Ga, Zn, Ag, and Ni), respectively~\cite{Takenaka2014}.
	
	After the ground-state charge density is obtained, a total of 80 maximally localized Wannier functions, including the $s$, $p$, and $d$ orbitals of Mn and $X$ atoms as well as the $s$ and $p$ orbitals of N atom, are disentangled from 144 Bloch bands on a uniform $k$-mesh of 10$\times$10$\times$10, using the \textsc{wannier90} package~\cite{Mostofi2008}.  Then, using the Kubo formula, intrinsic anomalous Hall conductivity [Eqs.~\eqref{eq:IAHC} and~\eqref{eq:BerryCur}] and optical conductivity [Eq.~\eqref{eq:optical}] are evaluated on a denser $k$-mesh of 100$\times$100$\times$100.  Finally, the Kerr [Eq.~\eqref{eq:kerr}] and Faraday [Eq.~\eqref{eq:faraday}] spectra are derived from the optical conductivity.


	%
	
\end{document}